# Partition of the total excitation energy between complementary fragments

*C. Manailescu[a], A. Tudora[a]*, F.-J. Hambsch[b], C. Morariu[a], S. Oberstedt[b]*

[a]*University of Bucharest, Faculty of Physics, Bucharest-Magurele, POB MG-11, R-76900, Romania*
[b]*EC-JRC Institute for Reference Materials and Measurements (IRMM), Retieseweg 111, 2440 Geel, Belgium*

**Abstract**
Two methods of the total excitation energy (TXE) partition between complementary fission fragments are compared. The first one is based on the "classical" hypothesis of prompt neutron emission from fully accelerated fission fragments with both fragments having the same residual nuclear temperature distribution. The second one is based on the systematic behaviour of the experimental multiplicity ratio $\nu_H/(\nu_L+\nu_H)$ as a function of the heavy fragment mass number $A_H$, the complementary fragments having different residual temperature distributions. The two TXE partition methods were applied to six fissioning systems: $^{233,235}$U($n_{th}$,f), $^{239}$Pu($n_{th}$,f), $^{237}$Np($n_{5.5MeV}$,f), $^{252}$Cf(SF), $^{248}$Cm(SF) and fragment excitation energies, level density parameters, fragment and fragment pair temperatures were compared. Limitations of the "classical" TXE partition method are pointed out. Residual temperature ratios $RT=T_L/T_H$ versus $A_H$ are obtained, as well as local and global parameterizations of RT($A_H$) for the neutron induced fissioning systems. Average values of quantities characterizing prompt neutron emission are discussed, too. A linear decrease of <RT> with the mass number of the fissioning nucleus and a linear decrease of the average C parameter with the fissility parameter is obtained. Point by Point model calculations are used to validate the RT($A_H$) parameterizations. The multi-parametric matrix ν(A,TKE) as well as prompt neutron and gamma-ray emission quantities as a function of fragment mass, total average prompt neutron multiplicity and spectrum and prompt neutron multiplicity distribution P(ν) were calculated. The results are in very good agreement with existing experimental data and evaluations. The global RT($A_H$) parameterization extends the use of the PbP model allowing the prediction of prompt neutron emission quantities for fissioning systems without experimental prompt neutron emission data. An explanation of the less pronounced sawtooth shape of ν(A) and the increase of ν(A) with incident neutron energy only for heavy fragments is given and exemplified by quantitative results of the Point by Point model.

*) Corresponding author
E-mail addresses: anabellatudora@hotmail.com, atudora@gmail.com

## 1. Introduction

As of today, the process of nuclear fission still represents a major challenge for the theoretical understanding as well as for experimental investigations. One of the long-standing question about the nuclear fission process is how does the available total excitation energy (TXE) gets partitioned between complementary light (LF) and heavy (HF) fragments. Several ideas and solutions were proposed during the past years in the frame of prompt neutron emission models (see for instance [1 - 9] and references therein).

In this work two methods of the TXE partition between complementary LF and HF are analyzed and compared:

***i)*** The first one is based on the "classical" hypothesis of prompt neutron emission from fully-accelerated fission fragments (FF) with both FF having the same residual nuclear temperature distribution. This is the hypothesis used in the "basic" Los Alamos (LA) model of Madland and Nix [1] and also in many papers devoted to the prompt neutron emission (see for instance Ref. [2]).

***ii)*** Taking into account that almost the entire prompt neutron emission takes place from fully-accelerated FF, the second method consists in the TXE partition between complementary FF in the same ratio as the numbers of prompt neutrons emitted by the LF and HF. This method leads to unequal residual temperatures of the complementary FF. This TXE partition has been used in the frame of the Point by Point (PbP) treatment based on parameterizations of the experimental sawtooth ν(A) data, (see Refs. [10 - 13] and references therein).

The TXE partition according to the first hypothesis of an equal residual nuclear temperature of fully-accelerated FF, usually used in LA type models, does not need experimental data of prompt neutron emission as a function of fragment mass. Excitation energies (E*) and the level density parameters "$a$" of FF are obtained simultaneously by an iterative procedure in the frame of the generalized super-fluid model of Ignatiuk [14]. This method was also applied in Ref. [2] using a Monte-Carlo approach of prompt neutron emission and considering a constant reaction cross-section of the inverse process.

In the following the TXE partition method of item i) will be labeled the "equal T" method.

Another TXE partition method with a non-equal residual temperature distribution of the FF was assumed in Ref. [2]. This approach was based on available experimental data for the prompt neutron sawtooth ($\nu_{\exp}(A)$), the average prompt neutron energy in the center-of-mass system ($<\varepsilon>_{\exp}(A)$) and the average prompt gamma-ray energy ($<E_\gamma>_{\exp}(A)$), both as a function of fragment mass to infer the initial excitation of each fragment. This method, based on solid physical considerations is useful for study purposes but is restrictive and with little predictive power because it requires too many experimental input data, available only for a few fissioning systems.

In the case of the TXE partition assuming different residual temperatures of complementary fully accelerated FF, the so-called fragment temperature ratio $RT = T_L / T_H$ can be introduced as a parameter. Originally, RT was defined by Oshawa [5] as the ratio of the average temperatures of the LF and HF in the frame of a simple LA model with only one fragmentation and the compound nucleus cross-section of the inverse process taken constant. In Ref. [5] the level density parameter $a$ is considered linear dependent on the mass number (with slope 1/C). This is an approximation that can be eventually used only at very high FF excitation energies, not appearing in spontaneous and neutron induced fission at moderate incident neutron energies.

Recently, more attempts were made to find average values of RT or functions describing this parameter. For instance Talou et al. [6, 7] took RT as a free parameter and proposed for $^{235}U(n_{th},f)$, $^{239}Pu(n_{th},f)$ and $^{252}Cf(SF)$ an average value <RT> = 1.2. For the $^{252}Cf(SF)$ case, Serot [9] gives a linear RT dependence on the LF mass number by taking into account the rotational and intrinsic energy components at the scission moment and using a Monte-Carlo treatment [8].

## 2 TXE partition methods in the frame of the Point by Point treatment

During the past years, the PbP model was applied to a large number of fissioning systems (including neutron induced and spontaneous fission) using both hypotheses of the TXE partition (see for instance Refs. [10-13] and references therein). This fact allowed us to deduce systematic behaviours of both prompt neutron experimental data and calculated quantities. Effects of the TXE partition methods on model parameters, i.e. the level density parameter *a*, the FF residual temperature and on prompt neutron emission quantities, i.e. the multi-parametric matrix ν(A,TKE), $\nu(A)$, could be investigated.

In the frame of the PbP model the TXE partition according to the ratio of prompt neutron numbers emitted by complementary FF was possible due to the following systematic behaviour deduced exclusively from experimental ν(A) data:

The experimental ν(A) data were represented as $\nu_H/(\nu_L+\nu_H)$ versus the HF mass number $A_H$ [10]. This representation was preferred over the traditional $\nu(A)$ because the nuclei forming the HF group do not change significantly from one fissioning system to another. We observed that for all fissioning systems having experimental *ν(A)* data, the quantity $\nu_H/\nu_{pair}$ as a function of $A_H$ exhibits a systematic behaviour with the following features:

- a minimum in $\nu_H/\nu_{pair}$ occurs around $A_H$=130 driven by the magic numbers Z=50, N=82 and the very high negative values of shell corrections,

- the complementary fragments emit almost an equal number of neutrons around the mass number 140 (the fragment pairs with $A_H$~140 being the most probable fragmentations),

- the LF emits more neutrons than the HF only in the range $A_H$ < 140 ($\nu_H/\nu_{pair}$ less than 0.5) while above $A_H$=140 the HF emits more neutrons than the LF ($\nu_H/\nu_{pair}$ great than 0.5).

The fact that $\nu_H \approx \nu_L$ at $A_H$~140 (where the HF mass yield distributions are at maximum, too) validates once again the assumption of Madland and Nix of "an equal number of neutrons emitted by the LF and HF" made in the case of the LA "most probable fragmentation" approach [1].

The TXE partition methods (items *i* and *ii* above) are compared to each other in the frame of the PbP treatment. Since experimental *ν(A)* data are very scarce, the 6 fissioning systems, with the best measured $\nu_{exp}(A)$ data have been investigated in the present work: three thermal neutron induced fissioning systems, $^{239}Pu(n_{th},f)$, $^{235,233}U(n_{th},f)$, one neutron induced fissioning system at higher incident neutron energy, $^{237}Np(n,f)$ at En = 5.5 MeV, and two spontaneously fissioning systems $^{252}Cf(SF)$ and $^{248}Cm(SF)$.

In all cases, according to the PbP treatment (see [10-13] and references therein), the FF range, was chosen as following: the entire fragment mass range covered by the experimental FF distributions with a step of 1 mass unit. For each mass unit 4 charge numbers Z are taken as the nearest integer values above and below the most probable charge obtained from the "unchanged charge distribution" corrected with a possible charge polarization.

For the six fissioning systems chosen in this study, the following experimental Y(A,TKE) distributions were used:

$^{239}$Pu($n_{th}$,f): Y(A) measured at IRMM [15] and TKE(A) of Wagemans et al. [16]
$^{235}$U($n_{th}$,f): Y(A,TKE) measured at IRMM [17]
$^{233}$U($n_{th}$,f): Y(A,TKE) of Surin et al. taken from EXFOR [18]
$^{237}$Np(n,f): Y(A,TKE) measured at IRMM [19]
$^{252}$Cf(SF): Y(A,TKE) measured at IRMM [20]
$^{248}$Cm(SF) Y(A,TKE) of PNPI (Vorobyev et al. [21])

A great part of the experimental FF distributions mentioned above were already used (see for instance Refs.[10 - 13], [22] and references therein) , with two exceptions: the experimental TKE(A) of Wagemans for $^{239}$Pu($n_{th}$,f) and the Y(A,TKE) distribution of Surin in the case of $^{233}$U($n_{th}$,f). Concerning the charge polarization, for $^{235}$U($n_{th}$,f) we used experimental ΔZ data fitted by Wahl, and taken from Ref. [23]. A detailed description of the PbP model can be found in Refs.[10 - 13] and references therein. We recall only that TXE of each pair of fragments is calculated as:

$$TXE(Z_L, A_L; Z_H, A_H) = E_r(Z_L, A_L; Z_H, A_H) + E_n + B_n(Z_{CN}, A_{CN}) - TKE(A_L, A_H) \qquad (1)$$

where $E_n$ is the incident neutron energy, $B_n(Z_{CN},A_{CN})$ is the neutron binding energy of the compound nucleus undergoing fission (for spontaneous fission both $E_n$ and $B_n$ are taken as zero). For the total kinetic energy distribution TKE(A) entering eq. (1) the experimental data mentioned above are used. The energy release in fission *Er* is calculated as the difference between the compound nucleus and FF mass excesses (taken from the database of Audi and Wapstra [24]):

$$E_r(Z_L, A_L; Z_H, A_H) = M(Z_{CN}, A_{CN}) - M(Z_L, A_L) - M(Z_H, A_H) \qquad (2)$$

The systematic behaviour of experimental $\nu_H/\nu_{pair}$ data versus $A_H$ mentioned above is illustrated in the upper part of **Figs.1a-f** for the 6 studied fissioning systems. The experimental data, taken from the EXFOR library [25] are plotted with different black and gray symbols. The solid lines are simple parameterizations (partly already reported in [10]). In all cases $\nu_H/\nu_{pair}$ shows a minimum at $A_H$=130 (except for $^{248}$Cm(SF)). In the case of the neutron induced fissioning systems an equal number of neutrons are emitted in the $A_H$ range of a few mass units around $A_H$=140, while for spontaneous fission (SF) this range is narrower (one-two mass units only) and shifted to $A_H$ above 140. It needs to be mentioned that the experimental Y(A) data of these SF systems [20, 21] have maxima at $A_H$ a few mass units above 140, too.

In the following the TXE partition method according to the hypothesis *ii)* will be labeled as the method based on the $\nu_H/\nu_{pair}$ parameterization.

### 3. Excitation energies and level density parameter *a* of fission fragments

Fission fragment excitation energies *E*(A)* and level density parameters *a(A)* of the 6 fissioning systems are calculated in the frame of the generalized super-fluid model [14] according to the two TXE partition methods, as given by the following eqs.:

$$a(E^*) = \begin{cases} \tilde{a}\left(1+\frac{\delta W}{U^*}(1-\exp(-\gamma U^*)\right) & U^* \geq U_{cr} \\ a_{cr} & U^* < U_{cr} \end{cases} \qquad U^* = E^* - E_{cond} \qquad (3)$$

with the condensation energy, $E_{cond} = \frac{3a_{cr}\Delta^2}{2\pi^2} - n\Delta$, where $\Delta = \frac{12}{\sqrt{A}}$ is the pairing correlation function and n = 0, 1 and 2 for even-even, odd-A and odd-odd nuclei, respectively. The critical temperature of the phase transition from super-fluid (super-conductive) to normal states is given by $t_{cr} = 0.567\Delta$ and the critical energy is $E_{cr} = a_{cr}t_{cr}^2$. The parameter of the function defining the damping of shell effects $\delta W$, is taken as $\gamma = 0.4A^{-1/3}$.

Eqs. (3) (referring to complementary LF and HF) are used in two ways: for the TXE method based on the $\nu_H/\nu_{pair}$ parameterization $E^*_{L,H}$ are calculated and entered in eq. (3) to provide the level density parameters $a_{L.H}$. For the "equal T" method, both quantities $E^*$ and $a$ are obtained simultaneously using eqs. (3) in an iterative procedure under the condition that both fragments have the same T value.

The *E*(A)* results are plotted in **Figs. 2a-f** with full squares in the case of the $\nu_H/\nu_{pair}$ parameterization method (given by the solid lines in the upper parts of Figs. 1a-f) and with full circles for the "equal T" method. The level density parameters *a(A)* are plotted using the same symbols in **Figs. 3a-f**. In the case of $^{252}$Cf(SF) (Figs. 2e, 3e) the open diamonds are *E*(A)* and *a(A)* obtained by applying eqs. (3) in the frame of an iterative procedure under the condition of the fragment temperature ratio $T_L/T_H = RT$ satisfying a linear dependence on the LF mass given by Serot [8, 9].

The thin dotted lines connecting the points in Figs. 2, 3 as well as in all the following figures are plotted only to guide the eye.

Looking at Figs. 2a-f a sawtooth-like behaviour of *E*(A)* is visible (and obviously expected in the case of the $\nu_H/\nu_{pair}$ parameterization). For the "equal T" method (full circles), the obtained *E*(A)* have less pronounced sawtooth shapes. For $^{252}$Cf(SF), as it can be seen in Fig. 2e, *E*(A)* values obtained from the RT function of Ref. [9] (open diamonds) are very close to the values using the $\nu_H/\nu_{pair}$ parameterization (full squares). Also the *E*(A)* result obtained from the present $\nu_H/\nu_{pair}$ parameterization is close to the $^{252}$Cf(SF) result of Ref. [2] obtained by using the experimental data for $\nu_{\exp}(A)$, $<\varepsilon>_{\exp}(A)$ and $<E_\gamma>_{\exp}(A)$, this fact being a verification of the hypothesis of method *ii)*.

The values of the level density parameter *a(A)* obtained using the two methods are close to each other, too, as it can be seen in Figs. 3a-f. For $^{252}$Cf(SF), the use of the RT function of Refs. [8, 9] gives similar results (see Fig. 3e).

The very close results for the level density parameter *a* obtained by the two TXE partition methods are more visible when *a* of the complementary FF is plotted as a function of $A_H$, an example being given in the upper part of **Fig. 4** for $^{235}$U($n_{th}$,f). The *a*-parameters obtained by the $\nu_H/\nu_{pair}$ parameterization are plotted with full squares and by the "equal T" method with open circles. Very good agreement is observed, only small differences being visible in the $A_H$ region around 130.

The ratios $a_H/(a_L+a_H)$ as a function of $A_H$ are plotted for comparison in the lower part of Figs. 1a-f, with full circles in the case of the "equal T" method and with open circles in the case of the $\nu_H/\nu_{pair}$ parameterization. As it can be seen the two methods lead to close results and for all fissioning systems the ratio $a_H/(a_L+a_H)$ exhibits a similar behaviour as the $\nu_H/\nu_{pair}$ ratio, with the observation that the minima in the $a_H/(a_L+a_H)$ ratios cover a broader mass range $A_H$=129-134.

In the case of *E*(A)* the less pronounced sawtooth shape obtained for the "equal T" method as well as the close *E** values provided by the two methods in the $A_H$ range 135-145 and especially for $A_H$≈140 are more visible if *E** is plotted in the same manner as the *a*-parameter focusing on the pair, see the lower part of Fig. 4.

The sawtooth-like behaviour of $\nu(A)$ and E*(A) as well as the fact that *E*(A)* of the "equal T" method has a less pronounced sawtooth shape can be explained by the energy conservation at the scission moment and the behaviour of the level density parameter *a* in connection with the shell-effects.

For a pair of fragments, the energy conservation at the moment of scission is given by ([3, 4] and references therein):

$$E_r + B_n + E_n = E_{pre} + E_{coul} + E_{def} + E_{\text{int}} \tag{4}$$

where $E_r$, $B_n$ and $E_n$ have the same meaning as in eqs. (1, 2). $E_{pre}$ and $E_{coul}$ are the pre-scission kinetic energy and the Coulomb repulsion energy between the two nascent fragments, respectively. $E_{def}$ is the sum of deformations energies of complementary fragments. The intrinsic energy (given by the dissipative and heating energies [3]) is shared between the complementary nascent fragments:

$$E_{\text{int}} = E_{dis} + E_h = E_{\text{int}}^L + E_{\text{int}}^H \tag{5}$$

After full acceleration of the fragments, the total kinetic energy is given by:

$$TKE = E_{pre} + E_{coul} \tag{6}$$

and the total excitation energy of the fully-accelerated FF becomes:

$$TXE = E_{def}^L + E_{def}^H + E_{\text{int}}^L + E_{\text{int}}^H = E_L^* + E_H^* \tag{7}$$

The intrinsic energy of the nascent fragments can be expressed as $E_{\text{int}}^{L,H} = a_{L.H}^{\text{int}}\,\tau_{L,H}^2$ and assuming statistical equilibrium at the scission moment [3, 4, 8] (equal nuclear temperatures $\tau_L = \tau_H$ of the nascent fragments), the intrinsic energy is shared between the nascent fragments in the same ratio as the level density parameters $a_{L,H}^{\text{int}}$:

$$\frac{E_{\text{int}}^L}{E_{\text{int}}^H} = \frac{a_L^{\text{int}}}{a_H^{\text{int}}} \tag{8}$$

At low and moderate energies the behaviour of the level density parameter *a* in different A and Z ranges is strongly influenced by the shell effects δW (according to eq. (3)). An example is given in the upper part of **Fig. 5** where the shell corrections (taken from Ref. [26]) of nuclei forming the FF range of $^{235}$U(n,f) are plotted without pairing corrections (full squares) and when the pairing corrections of Ref. [27] are taken into account (stars). As it can be seen, for HF with 125<$A_H$<140 the shell corrections show very high negative values (due to magic and doubly magic nuclei with Z=50, N=82). The corresponding light fragments (111>$A_L$>96) have almost constant positive shell corrections.

For the same FF range, the behaviour of the level density parameter *a* with the excitation energy is illustrated in the lower part of Fig. 5 for two cases: for excitation energy values of 2 MeV (open squares) and of 25 MeV (full circles). As it can be seen the level density parameter *a* exhibits a very pronounced increase with the excitation energy around masses 130, 132, while in the range 90-120 it slowly decreases with increasing excitation energy. For far asymmetric fragmentations (HF with $A_H$ above 145 and LF with $A_L$ less than 80) the level density parameter does not vary much with excitation energy.

For fragment pairs with $A_H$ around 140 (the most probable fragmentation) the *a*-values of complementary fragments are very close at any excitation energy. In **Fig. 6** an example of the *a*-parameter variation with excitation energy for the most probable fragmentation ${}^{140}_{54}Xe,\ {}^{96}_{38}Sr$ is given with full and open up triangles, respectively. Consequently, for pairs with $A_H$≈140, the intrinsic energy is shared in almost equal parts between the two fragments. For fragment pairs with $A_H$>140 the *a*-values for HF are higher than for LF at any excitation energy. See for instance in Fig. 6 the variation of *a* with the energy for the fragmentation ${}^{145}_{56}Ba,\ {}^{91}_{36}Kr$ plotted with full and open down triangles. Consequently, for pairs with $A_H$>140 $E^H_{\text{int}} > E^L_{\text{int}}$. The HF with $A_H$>140 are pronounced deformed nuclei while the complementary LF are spherical or less deformed, hence more deformation energy being stored into the HF. As a consequence for pairs with $A_H$>=140 the excitation energies of the complementary fully-accelerated fragments (as a sum of deformation and intrinsic energies) fulfill the condition $E^*_H \geq E^*_L$, this fact being confirmed by all experimental ν(A) data showing $\nu_H \geq \nu_L$ for pairs with $A_H \geq 140$ (see the upper parts of Figs. 1a-f).

For pairs with $A_H$<140 the values of *a* for LF are higher than for HF at any excitation energy (see Fig. 5, lower part). Also an example is given in Fig. 6 for the fragmentation ${}^{125}_{49}In,\ {}^{111}_{43}Tc$ (with full and open circles). Especially in the $A_H$ range around 132 the $a_H$-values are very low compared with complementary $a_L$ values, see for instance in Fig. 6 the variation of *a* with the energy for the fragmentations ${}^{130}_{50}Sn,\ {}^{106}_{42}Mo$ (full and open squares) and ${}^{134}_{52}Te,\ {}^{102}_{40}Zr$ (full and open diamonds). Consequently, for pairs with $A_H$<140 much more intrinsic energy is stored into the LF ($E^L_{\text{int}} > E^H_{\text{int}}$). The nuclei with $A_H$ around 130-132 are almost spherical (due to magic numbers Z=50, N=82) and practically the entire deformation energy is found in the complementary light fragment. As a consequence for pairs with $A_H$<140 the excitation energies of fully accelerated FF fulfill the condition $E^*_L > E^*_H$ with $E^*_L >> E^*_H$ for fragment pairs with $A_H$ around 130-132. This fact is also proved by the behaviour of experimental ν(A) data (see the upper parts of Fig. 1) exhibiting $\nu_L > \nu_H$ for pairs with $A_H$<140 and a minimum of $\nu_H$ for pairs with $A_H$≈130.

In conclusion, when the "equal T" method is used, then the total excitation energy TXE of the complementary fully-accelerated FF is partitioned by the ratio $a_L/a_H$ and not the intrinsic excitation energy of the nascent fragments. The neglection of the deformation energy contribution leads to a less pronounced sawtooth shape of E*(A) compared to the TXE partition according to the multiplicity ratio.

The close values of $a$ provided by the two TXE partition methods in almost the entire mass range (exceptions only around the mass number 130-132) is due to the slow variation of the level density parameter $a$ with energy (illustrated in the lower part of Fig. 5) compared with the pronounced dependence of $a$ on the shell correction.

The $C$-parameter of the LA model [1] can be calculated for each FF pair as:

$$C = \frac{A_{CN}}{a_L + a_H} \quad . \qquad (9)$$

For the "equal T" method it has the same physical significance as in the "classical" LA model [1, 2]. In the case of the $\nu_H / \nu_{pair}$ parameterization (obviously leading to different residual temperature values of the LF and HF forming a pair) the close values of $a$ obtained by the two methods make the determination of an "equivalent" $C$-parameter possible (according to eq. (9)). The $C$-parameters obtained by the two methods are also close to each other, an example being given in **Fig. 7** for $^{239}$Pu(n$_{th}$,f) and $^{235}$U(n$_{th}$,f). As it can be seen the $C$-values are far from a constant value representing the slope of the $a$-parameter in the rough linear approximation $a = A/C$ (used in old prompt neutron emission models). An almost linear dependence on A of the $a$-parameter (with slopes of about 1/8 – 1/9MeV$^{-1}$) is obtained only at very high $E$* (around 100 MeV, not applicable for SF and neutron induced fission discussed here). Then the values of $a$ given by the super-fluid model tend to approach the asymptotic value $\tilde{a}$. And the usual parameterizations of $\tilde{a}$ [14, 28] are rather close to linear dependences on A.

### 4. Maximum value of the residual temperature distribution of fission fragments and fragment residual temperature ratios

The maximum residual temperatures of fragments are plotted in **Figs. 8a-f**, with full circles in the case of the $\nu_H / \nu_{pair}$ parameterization and with stars for the "equal T" method.

As expected the fragment temperatures (full circles) are practically equal to the temperature values of the "equal T" method (stars) in the fragment mass regions $A_H \approx$ 135-155 and $A_L \approx$ 85-105 because the $E$* values obtained by the two methods are close to each other in these mass regions and the values of $a$ of the two methods are close to each other almost in the entire A range. Taking into account that the mass distributions Y(A) show the highest yields in the mass ranges mentioned above, average values of prompt neutron emission quantities provided by the two methods are expected to be close to each other, too.

The close values of $a$ obtained by the two methods allow the introduction of an "equivalent" temperature of the FF pair as follows:

$$TXE = a_L^* T_L^2 + a_H^* T_H^2 = (a_L^* + a_H^*) T_{equiv}^2 \qquad (10)$$

where $a_{L,H}^*$ are the level density parameters obtained in the case of the $\nu_H / \nu_{pair}$ parameterization method of the TXE partition.

The obtained equivalent temperatures, plotted with open squares in Figs. 8a-f, are very close to the temperatures obtained from the “equal T” method (stars). In the $^{252}$Cf(SF) case, the equivalent T obtained from the RT function of Serot [9], plotted with open diamonds in Fig. 8e, are also very close to the equivalent T values obtained from the $\nu_H / \nu_{pair}$ parameterization (open squares) and to the equal T values (stars).

To highlight the proximity of the values, the ratios between the equal T and equivalent T of the six studied fissioning systems are plotted together in **Fig. 9**. As it can be seen, these ratios are practically 1 in almost the entire $A_H$ range, only around $A_H$=130 the maximum differences between the equivalent T and equal T values are about 4% in the case of neutron induced fissioning systems and about 6% in the case of spontaneous fission.

The fact that the residual temperatures T (of the “equal T” method) and the equivalent T (of the $\nu_H/\nu_{pair}$ parameterization method) are practically equal has as consequence that the LA type models (where T is the maximum value of the residual temperature distribution) are not sensitive to the TXE partition especially when only average prompt neutron emission quantities are considered.

The ratios between the maximum residual temperature of the LF and HF forming a pair are plotted in **Fig. 10** as a function of $A_H$ with different symbols for the studied neutron induced fissioning systems. A very interesting behaviour of the temperature ratio RT($A_H$) is visible and has the following features: a) the maximum of RT($A_H$) occurs at $A_H$=130 and is around 1.5-1.6, b) in the $A_H$ range between 135-145 the temperature ratio is approximately 1 (HF and LF having practically the same residual temperature) and c) for $A_H$>145 the decrease of RT is almost linear and the slope does not differ very much from one neutron induced fissioning system to another. This systematic trend allows for parameterizations of RT as a function of $A_H$, which are plotted with different line styles in Fig. 10, too. **Fig. 11** shows the slopes and intercepts of the RT parameterizations plotted versus the fissility parameter. Their almost constant values suggest that an unique RT parameterization for the neutron-induced fissioning systems can be deduced, given by the dashed lines in Fig. 11.

Temperature ratios of the two studied SF systems as a function of $A_H$ are plotted in **Fig. 12** and their behaviour differs from the RT behaviour of (n,f) as follows: a) the maximum of RT (around $A_H$=130) is higher than in the case of (n,f), b) the $A_H$ range where RT is 1 is limited to one-two mass units around $A_H$=145. Both SF systems have practically the same RT values in the $A_H$ range above 134. The visible differences in the region $A_H$<134 are mainly due to the shifted minimum of the experimental $\nu_H/\nu_{pair}$ data in the case of $^{248}$Cm(SF). Similar RT parameterizations as mentioned before have been obtained for the two SF systems, see the lines and the analytical expressions given in Fig. 10.

In Figs. 10 and 12 also <RT> values averaged over the FF mass and charge distributions are given. In the case of (n,f) <RT> are close to 1 , in the case of SF <RT> are about 1.1 to 1.15.

## 5. Examples of PbP calculation of prompt neutron emission quantities using the TXE partition methods and the new RT parameterizations

Almost all fissioning systems having experimental FF distributions were studied in the frame of the PbP model ($^{232,233,235,238}$U(n,f), $^{231,233}$Pa(n,f), $^{237}$Np(n,f), $^{239}$Pu(n,f), $^{240,242}$Pu(SF), $^{244,248}$Cm(SF), $^{252}$Cf(SF)) with results reported in [10-13] and references therein. For most of these fissioning systems, the TXE partition into the ratio $\nu_L/\nu_H$ (based on the $\nu_H/\nu_{pair}$ parameterization) was successfully used, providing:
- the multi-parametric matrix ν(A,TKE) allowing to obtain many quantities related to each fragment (such as *ν(A)*, <ε>(A), $E_\gamma(A)$ and so on)
- average quantities (total average prompt neutron multiplicity and spectrum, <ν>(TKE) and so on)
- the prompt neutron multiplicity distribution P(ν).

For a few of the studied fissioning systems mentioned above (like $^{240,242}$Pu(SF) [13]) because experimental ν(A) data are missing, the "equal T" method has been used, leading to P(ν) results describing very well the experimental P(ν) data (see details in Ref. [13]).

For all fissioning systems the PbP calculations of *ν(A)* by using the two TXE partition methods showed that in the case of the "equal T" method a less pronounced sawtooth shape of *ν(A)* is obtained, due to the fact that in this method the deformation energy contribution is neglected and TXE is partitioned by the ratio $a_L/a_H$ instead of the intrinsic energy. An example is given in **Fig. 13**.

The calculated total prompt neutron spectrum based on the two TXE partition methods lead to very close results insignificant differences are only visible in the region of high prompt emitted neutron energies. An example is given in **Fig. 14** for $^{233}$U($n_{th}$,f), where the result obtained using the new RT parameterization is also plotted (with a dotted line).

However, both methods give a harder spectrum and a better description of the experimental data [29] compared to the ENDF/B-VII evaluation [30].

Even if the ν(A) sawtooth shapes provided by the two TXE partition methods are different, the total average prompt neutron multiplicity is practically insensitive to the TXE partition (two examples are given in Table 1).

**Table 1:**
Total average prompt neutron multiplicity from the two TXE partition methods

| TXE partition method | $^{239}$Pu($n_{th}$,f) | $^{237}$Np($n_{5.5MeV}$,f) |
|---|---|---|
| $\nu_H/\nu_{pair}$ parameterization | 2.8678 | 3.4417 |
| Equal T method | 2.8686 | 3.4437 |

This fact is due to the very close *ν(A)* values given by the two TXE partition methods in the fragment mass ranges where the FF mass distributions Y(A) show the highest yields. Since this results in very close FF residual temperatures and excitation energies (see for instance in Figs. 2 and 8 the very close values of *E*(A)* and *T(A)* in the mass ranges 90-105 and 135-150), total average prompt neutron and gamma-ray quantities insensitive to the TXE partition are expected.

The existing experimental ν(A) data for a given fissioning system show that the sawtooth shape of ν(A) is diminished when the incident energy is increasing, exhibiting a visible increase only of the multiplicity of HF (see for instance Ref. [31]). This experimental observation is rather general and it was also found for protons as incident particles on actinides like $^{233,238}$U (details can be found in Refs. [32-34]). In the case of $^{237}$Np(n,f), the ν(A) data measured at two incident energies 0.8 and 5.55 MeV [35], also show a visible multiplicity increase only for heavy fragments, see the upper part of **Fig. 15**.

This ν(A) behaviour can be easily explained by the pronounced and fast increase of the intrinsic energy of the HF for pairs with $A_H$<140 and especially for pairs with $A_H$ around 130-132, (according to eq. (8) and the behavior of the level density parameter *a* in Fig. 6 and the lower part of Fig. 5). This pronounced increase of the intrinsic energy for pairs with $A_H$<140 plays the most important role in the increase of total E* of HF, having as consequence the visible increase of $\nu_H$.

For pairs with $A_H$ above 140 (for which $\nu_H > \nu_L$ at any incident energy) and especially for far asymmetric pairs (i.e. with $A_H \geq 150$) the intrinsic energy ratio between HF and LF does not change significantly with increase of the incident energy (because of the very slow variation of *a* with energy). The fragments with $A_H$>150 are deformed nuclei and with the increasing incident energy more and more deformation energy is found into these HF. Consequently for pairs with $A_H$>150, almost the entire increase of the incident energy is found in the total E* of HF, reflected by an increase of the multiplicity only for HF.

Our PbP model calculations of ν(A) at the incident energies of 0.8 MeV (full circles) and 5.5 MeV (stars), describing well the scattered experimental data of Ref. [35], confirm the observed multiplicity increase with incident energy only for HF, as it can be seen in the lower part of **Fig. 15**.

### 6. Average values of quantities related to the TXE partition between FF

In many papers only the LA "most probable fragmentation" approach requiring average model parameters is used. Hence, determination of average values of quantities related to the TXE partition methods, part of them being also input parameters for models, is useful and leads to interesting conclusions.

All discussed quantities were averaged over the experimental FF mass distributions of Refs. [15-21] and charge distributions taken as narrow Gaussians according to Ref. [23].

The temperature ratios RT($A_H$) (plotted in Figs. 10 and 12) were averaged over Y(A) according to:

$$< RT >= \frac{\sum_{i=1}^{NF/2} RT_i \, Y(A_H^i)}{\sum_{i=1}^{NF/2} Y(A_H^i)} \quad , \quad RT_i = \frac{T_L^i}{T_H^i} \qquad \text{(NF = number of FF)} \qquad (11)$$

with the $RT_i$ values at each mass number A already averaged over the fragment charge distribution (4 Z values being taken at each A). The <RT> results according to eq. (11) are given in Figs. 10 and 12, too.

The average values of the C-parameter were calculated as:

$$< C >= \frac{A_{CN}}{< a >} \qquad \text{with } < a >= \frac{\sum_{i=1}^{NF/2} (a_L^i + a_H^i) Y_i}{\sum_{i=1}^{NF/2} Y_i} \qquad (12)$$

were the index i is running over the fragment mass pair range. Using 4 Z values per A, the level density parameters $a_{L,H}^i$ entering eq. (12) were already averaged over the FF charge distribution. The obtained average *C* parameter shows a linear decrease with the fissility parameter, as it can be seen in **Fig. 16**.

The average $C$-parameters of all fissioning nuclei studied during the last years are obtained around the value of $<C> = 11\ MeV$ given by Madland and Nix [1] (for details see the systematic behaviour of average input parameters of the LA model in Ref. [36]).

Average temperature ratios can be obtained in three ways:

a) by averaging the temperature ratio of each FF pair over the FF mass and charge distributions (as mentioned above, eq. (11));

b) by calculating the average residual temperatures of the LF and HF groups $<T_{L,H}>$ (also by averaging over the FF mass and charge distributions) and defining the temperature ratio as $<T_L>/<T_H>$;

c) by calculating the average excitation energies $<E^*_{L,H}>$ and level density parameters $<a_{L,H}>$ of the LF and HF groups, the mean residual temperatures of the LF and HF groups being obtained by the relation: $\overline{T}_{L,H} = \sqrt{<E^*_{L,H}>/<a_{L,H}>}$.

For all fissioning systems studied in this work the temperature ratios obtained in the three ways mentioned above, fulfil the relation:

$$\frac{<T_L>}{<T_H>} < \langle RT \rangle < \frac{\overline{T}_L}{\overline{T}_H} \qquad (13)$$

A linear decrease of the average temperature ratios as a function of the mass number of the fissioning nucleus is observed in the case of neutron induced fission, see **Fig. 17**.

For all studied fissioning systems and for both TXE partition methods the average excitation energies of the LF group are higher than the average $E^*$ of the HF group.

The average multiplicity of the LF and HF groups obtained from the PbP calculated sawtooth agrees with the average multiplicities obtained from the experimental sawtooth data [25] and confirms the fact that –only on average- the LF group emits more prompt neutrons than the HF group at low incident neutron energy.

## 7. Conclusions

The comparative analysis of the two TXE partition methods in the frame of the PbP treatment can be synthesized as follows:

- The systematic behaviour of experimental data concerning $\nu_H/\nu_{pair}$ as a function of $A_H$ leads to parameterizations that are used for the TXE partition between fully accelerated FF.

- The fragment pair residual temperature ratios RT as a function of $A_H$ can be linearly parameterized. The deduced RT($A_H$) parameterizations were verified by the PbP model calculations of quantities characterizing prompt neutron emission. For neutron induced fissioning systems a general parameterization of RT($A_H$) is proposed allowing the use of the PbP model to predict prompt neutron emission quantities of fragments.

- The maximum residual temperature ratios are practically equal to 1 for fragment pairs with $A_H$ in the range of a few mass units above and below 140 (where Y(A) are maximum, too). Consequently the obtained average RT values are only a little bit higher than 1. In other words for the fragmentations occurring with high probability the fragments have almost the same residual temperature distribution. For this reason when LA models with only one fragmentation (the so-called most probable fragmentation) are used, then the "classical" hypothesis [1] of the equal residual temperature distributions is re-confirmed and also recommended.

- The level density parameter ratios $a_H/(a_L+a_H)$ as a function of $A_H$ exhibit a similar behaviour as the experimental $\nu_H/\nu_{pair}$ ratios, only the minimum is less pronounced and placed diffusely between $A_H$=129-134. This leads to a less pronounced sawtooth shape of *E*(A)* and *ν(A)* in the case of the "equal T" method. This is due to the fact that in the case of the "equal T" method *TXE is partitioned by the ratio* $a_L/a_H$ *instead of the intrinsic energy of the nascent fragments, the deformation energy contribution being neglected*.

- The level density parameter values *a(A)* obtained using the two TXE partition methods are close to each other almost in the entire FF mass range, except the region around A=130 and $A_{CN}$-130. This is due to the slow variation of *a* with excitation energy in comparison with the pronounced influence of shell effects in different A and Z regions of the FF range. The most important variation of *a* with energy is in the mass region of nuclei with very high negative values of shell corrections (see Fig. 5).

As an immediate consequence of close values of *a* provided by the two TXE partition methods, the *C*-parameters are close to each other, too. The nearly equal values of *a(A)* of the two methods allow to define an "equivalent temperature" of the FF pair with values practically equal to the temperatures of the "equal T" method.

- The average temperature ratios defined as the ratio of average residual temperature of the LF and HF groups $<T_L>/<T_H>$, exceed 1 with no more that 6% in the case of neutron induced fission and with no more than 15% in the case of SF, being significantly lower than the value of 1.2 proposed in Ref. [7].

- The use of the linear RT function of [8, 9] available in case of $^{252}$Cf(SF), leads to values of prompt neutron quantities very close to the present results even if a higher average RT of about 1.3 is obtained compared to present TXE partition method based on the $\nu_H/\nu_{pair}$ parameterization. This fact proves that the model has no significant sensitivity to RT especially when total average quantities are calculated.

- The usual statement that at low excitation energies of the fissioning systems "the light fragments emit more neutrons than the heavy fragments" is true and verified only in the case of average multiplicities of LF and HF groups. But a more attentive analysis of experimental sawtooth data shows that the LF emits more neutrons than the HF only for fragment pairs with $A_H$ less than 140, for pairs with $A_H$ above 140 the HF emits more neutrons.

- Both TXE partition methods lead to average excitation energies $<E^*_L>$ of the LF group higher than $<E^*_H>$ of the HF group.

- Average values of the *C*-parameter (using both TXE partition methods) show a slow linear decrease with the fissility parameter.

- In the case of neutron-induced fissioning systems, the average RT values exhibit a linear decrease with the mass number of the fissioning nucleus.

Taking into account that almost the entire prompt neutron emission takes place from fully accelerated FF, the TXE partition between complementary fragments in the same ratio as the prompt neutron numbers (based on the $\nu_H/\nu_{pair}$ parameterization) can be considered the correct method.

At the scission moment statistical equilibrium is assumed for the nascent fragments, leading to an intrinsic excitation energy partition based on the level density parameter ratio $a_L/a_H$. For this reason the partition of TXE (as a sum of intrinsic and deformation energies) according to the ratio $a_L/a_H$ can be considered as an approximation because the contribution of the deformation energy of fragments is neglected.

The rather close results provided by the two TXE partition methods (especially when average quantities are concerned) proves that in the TXE partition the intrinsic energy partition plays a much more important role than the deformation energy component.

The sawtooth shape of *ν(A)* as well as the *ν(A)* behaviour with increasing incident neutron energy consisting in the diminution of the sawtooth shape and the multiplicity increase only for heavy fragments are consistently explained by the level density parameter of the generalized super-fluid model and the statistical equilibrium assumption of the complementary nascent fragments at the scission moment, being supported by the quantitative results of the PbP model.

**Acknowledgements**

One of us (A.T.) wants to acknowledge the support of the EFNUDAT project (Contract No. 036434). Also a part of this work (referring to the prompt neutron emission of $^{239}Pu(n_{th},f)$ and $^{233}U(n_{th},f)$) was done in the frame of the IAEA Research Contract No. 15805.

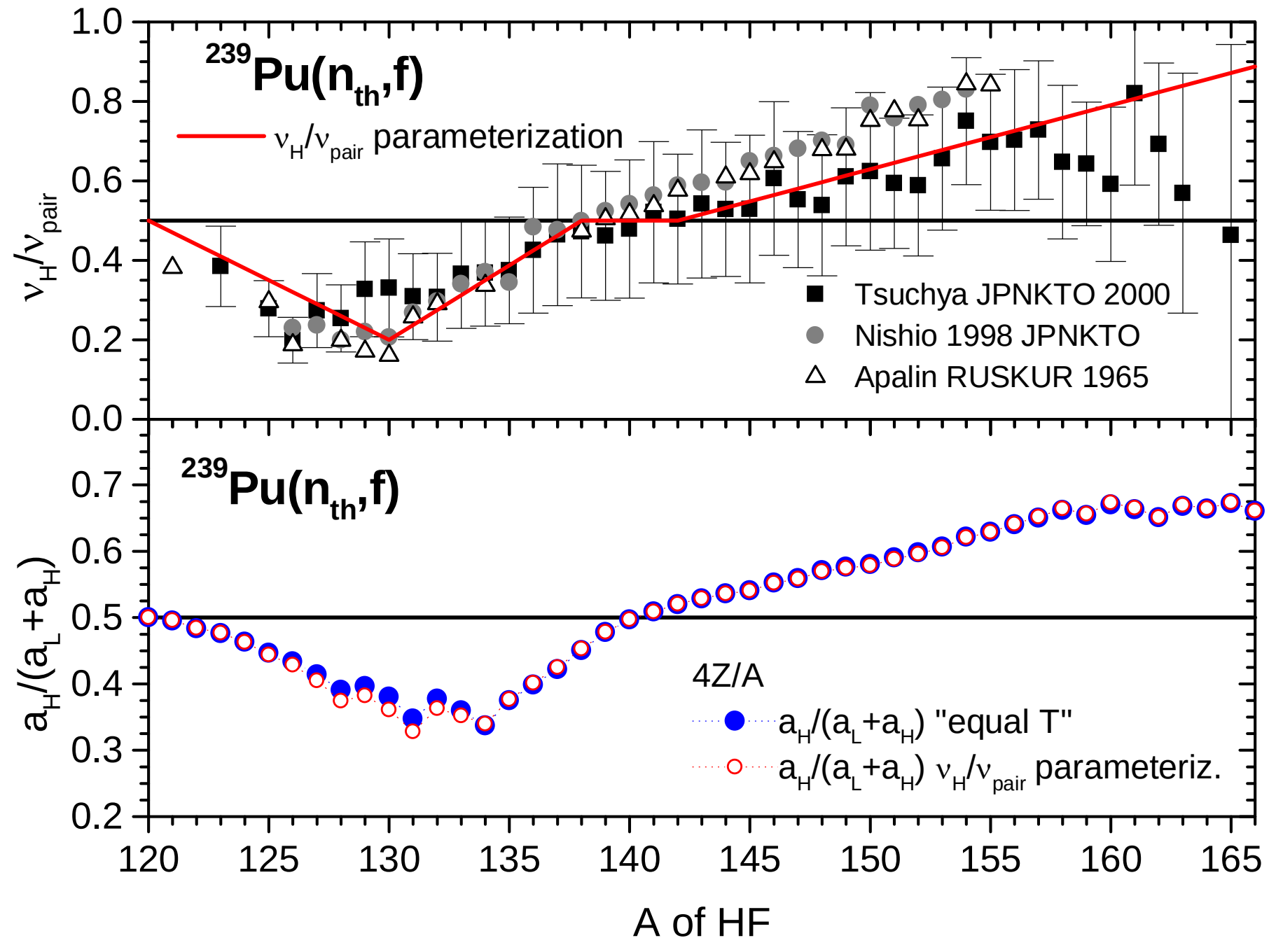

**Fig 1a**: $^{239}$Pu($n_{th}$,f)) upper part: $\nu_H/\nu_{pair}$ parameterization in comparison with experimental data, lower part the ratio $a_H/a_{pair}$ obtained in the two cases.

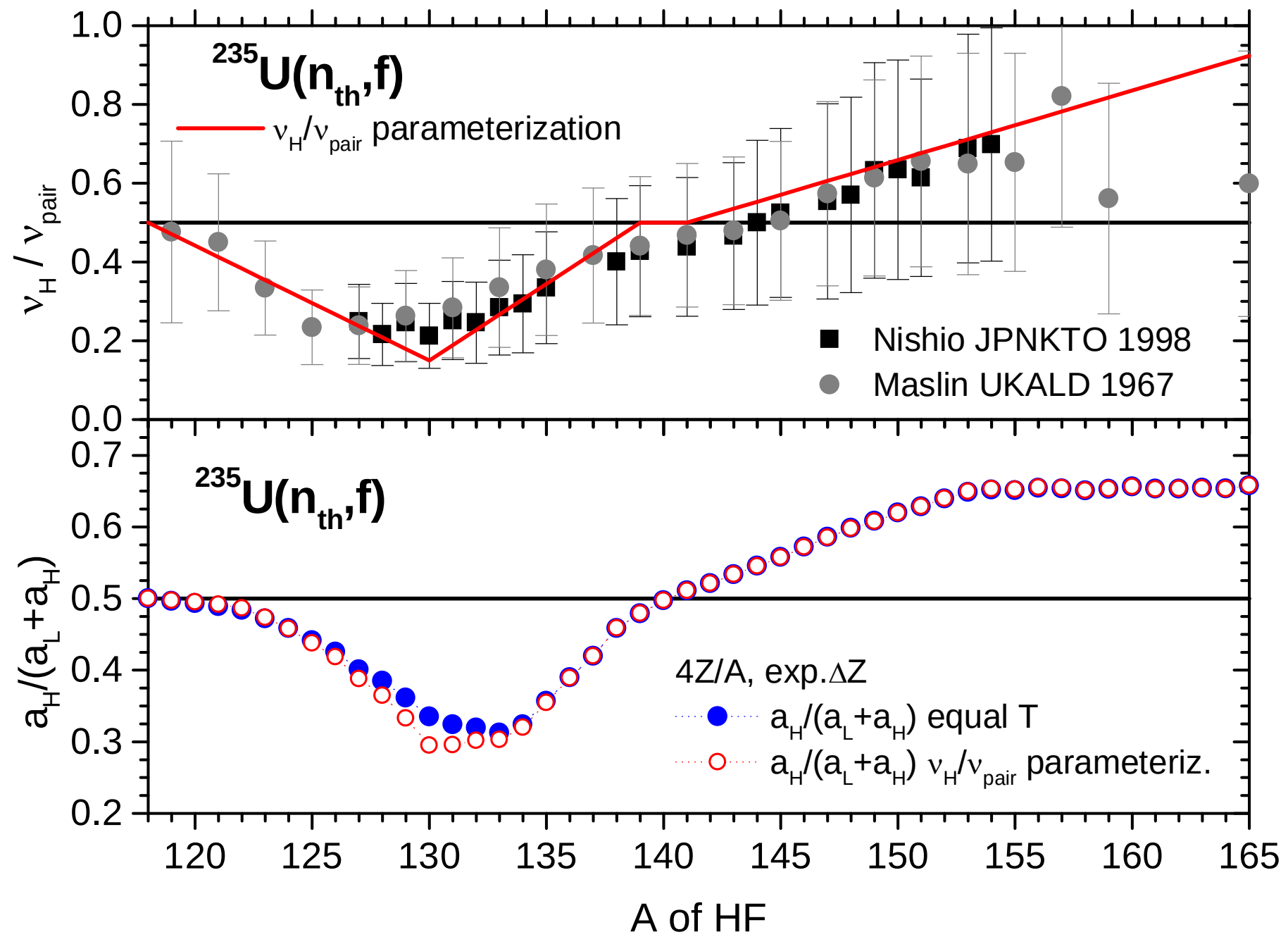

**Fig 1b**: $^{235}$U($n_{th}$,f)) upper part: $\nu_H/\nu_{pair}$ parameterization in comparison with experimental data, lower part the ratio $a_H/a_{pair}$ obtained in the two cases.

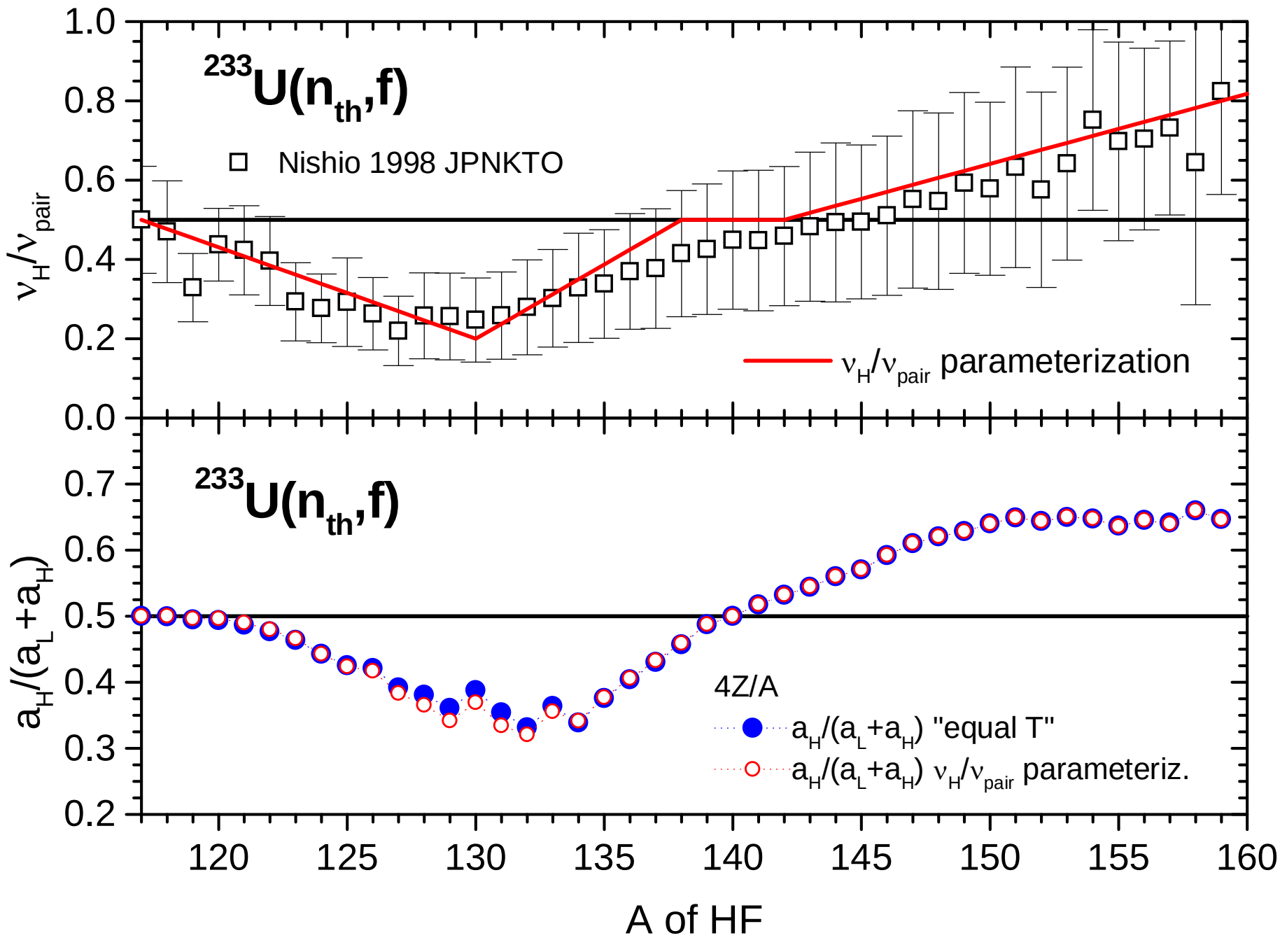

**Fig 1c**: $^{233}$U($n_{th}$,f)) upper part: $\nu_H/\nu_{pair}$ parameterization in comparison with experimental data, lower part the ratio $a_H/a_{pair}$ obtained in the two cases.

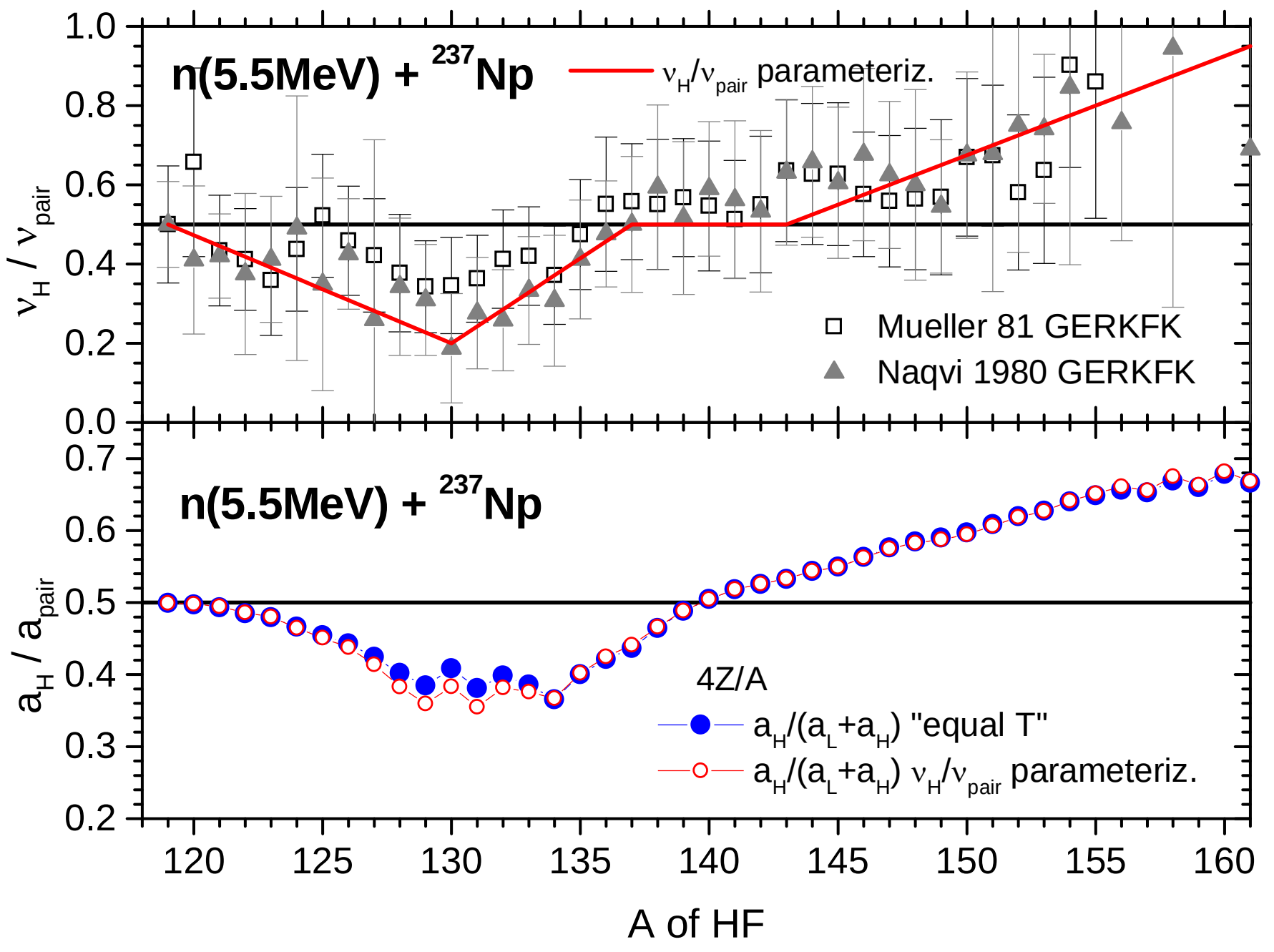

**Fig 1d**: $^{237}$Np(n,f) at En = 5.5 MeV, upper part: $\nu_H/\nu_{pair}$ parameterization in comparison with experimental data, lower part the ratio $a_H/a_{pair}$ obtained in the two cases.

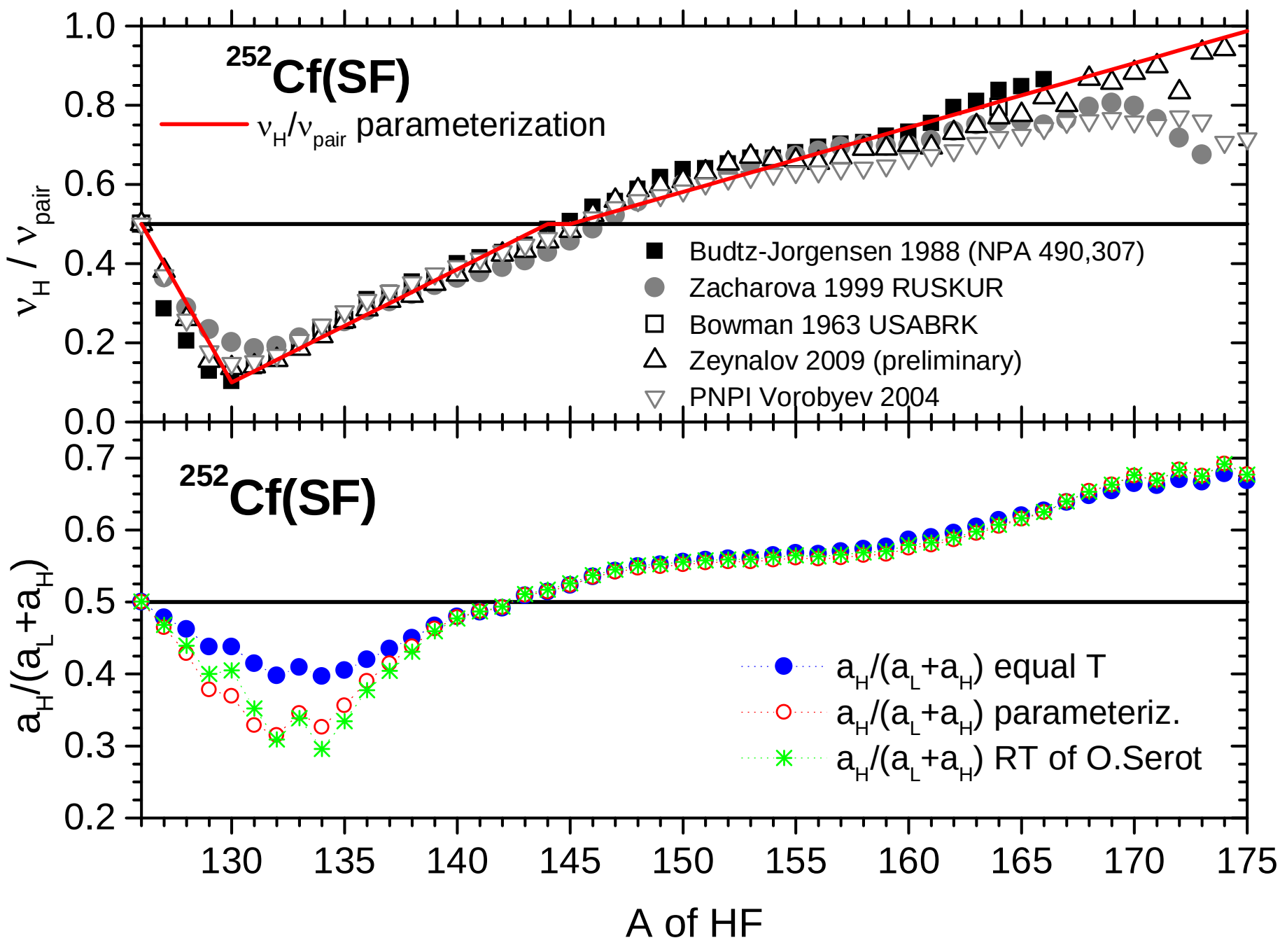

**Fig 1e**: $^{252}$Cf(SF)Upper part: $\nu_H/\nu_{pair}$ parameterization in comparison with experimental data, lower part the ratio $a_H/a_{pair}$ obtained in the three studied cases.

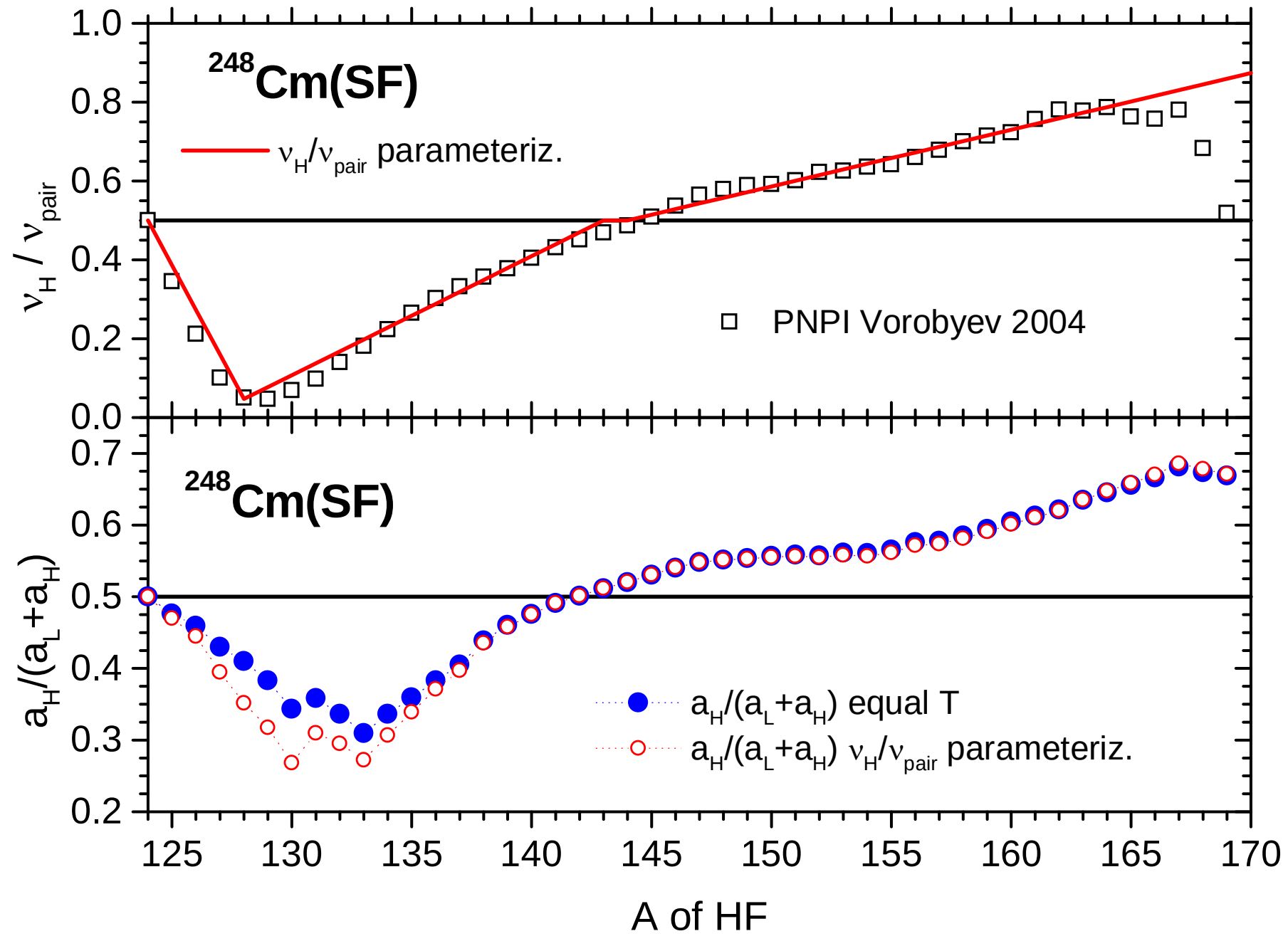

**Fig 1f**: $^{248}$Cm(SF) upper part: $\nu_H/\nu_{pair}$ parameterization in comparison with experimental data, lower part the ratio $a_H/a_{pair}$ obtained in the two cases.

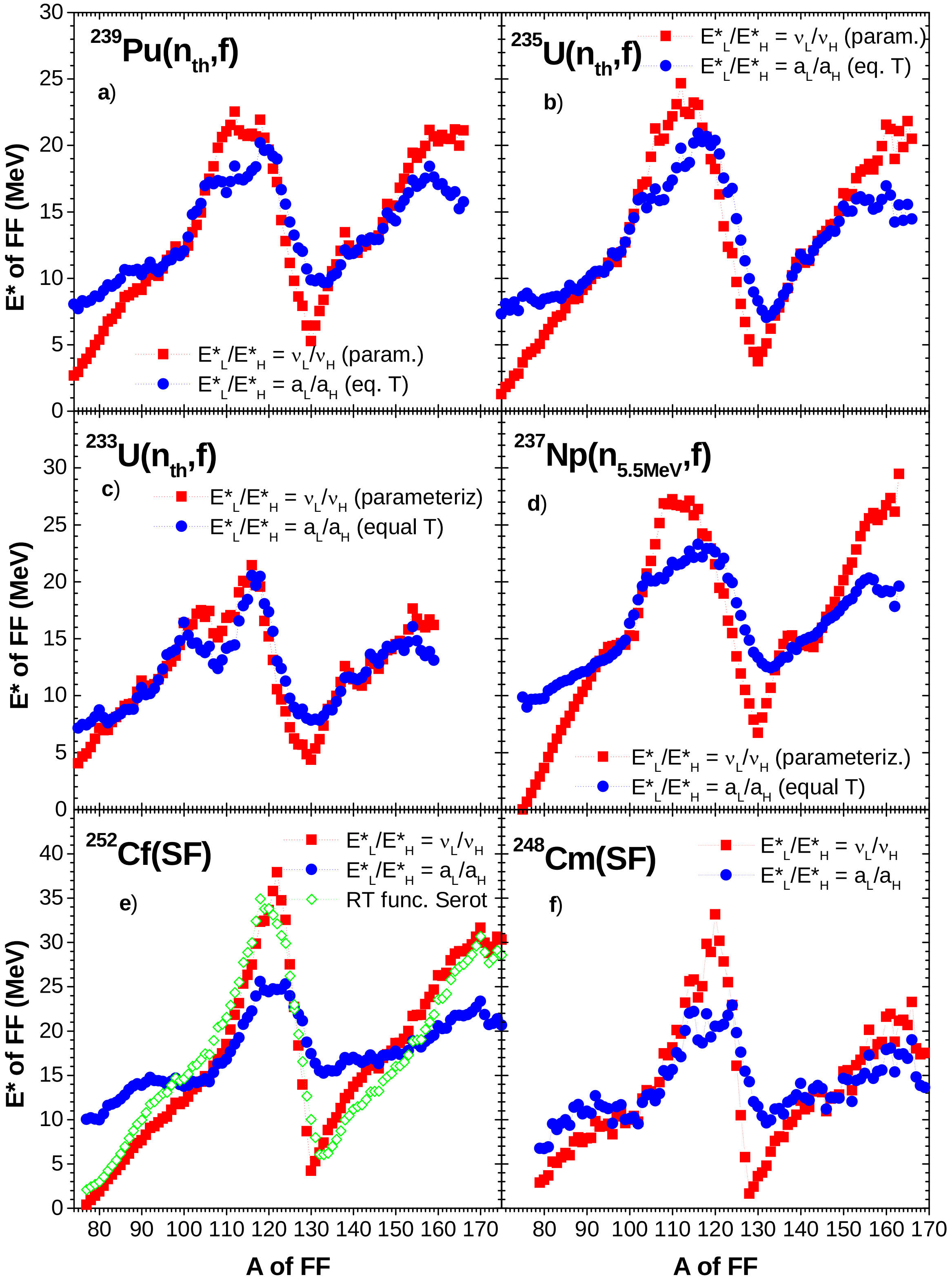

**Fig.2**: E*of FF obtained from the $\nu_H/\nu_{pair}$ parameterization and from the "equal T" method (iterative procedure).

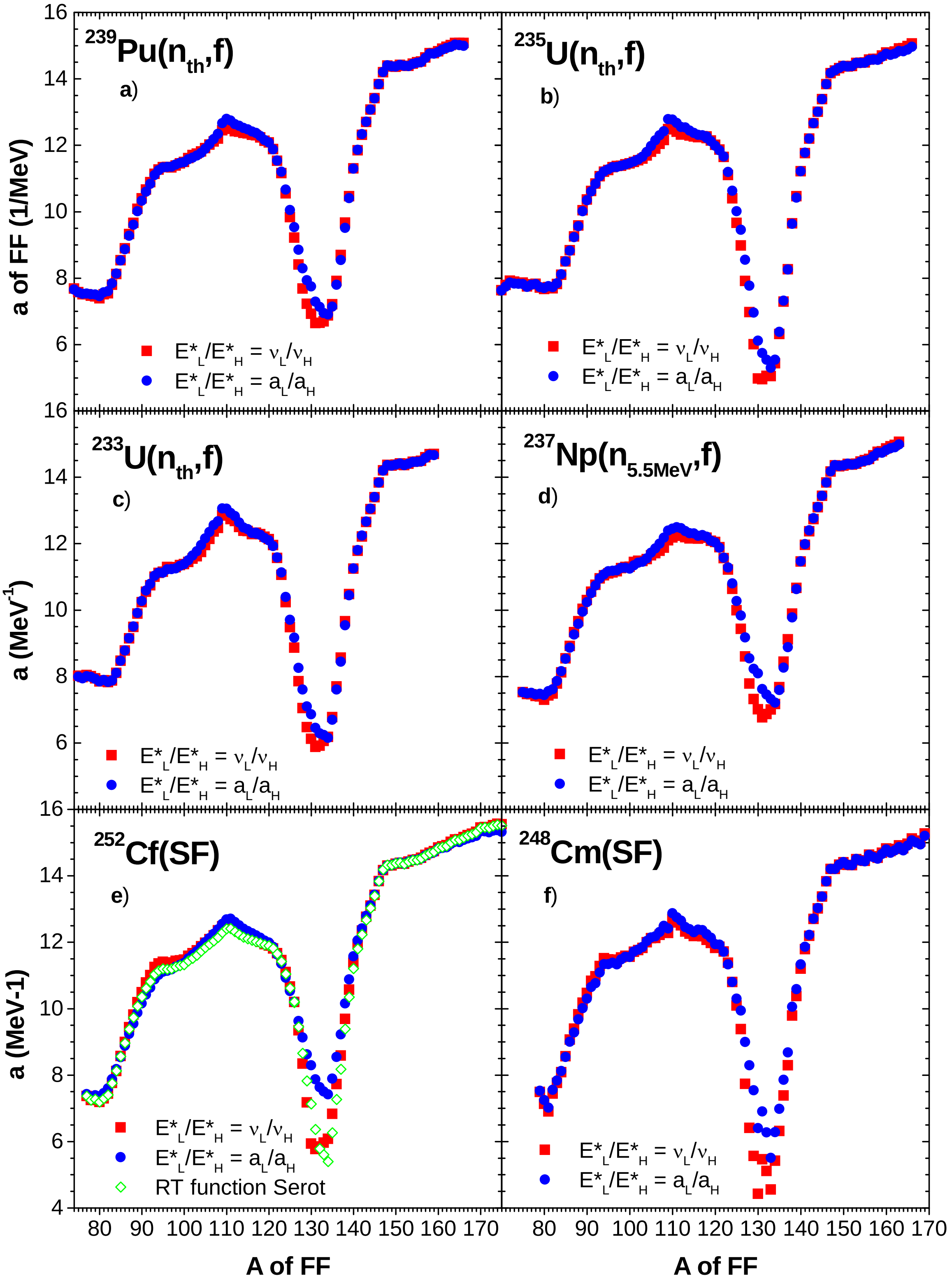

**Fig.3**: Level density parameter (the super-fluid model) obtained from the $\nu_H/\nu_{pair}$ parameterization and from the "equal T" method (iterative procedure).

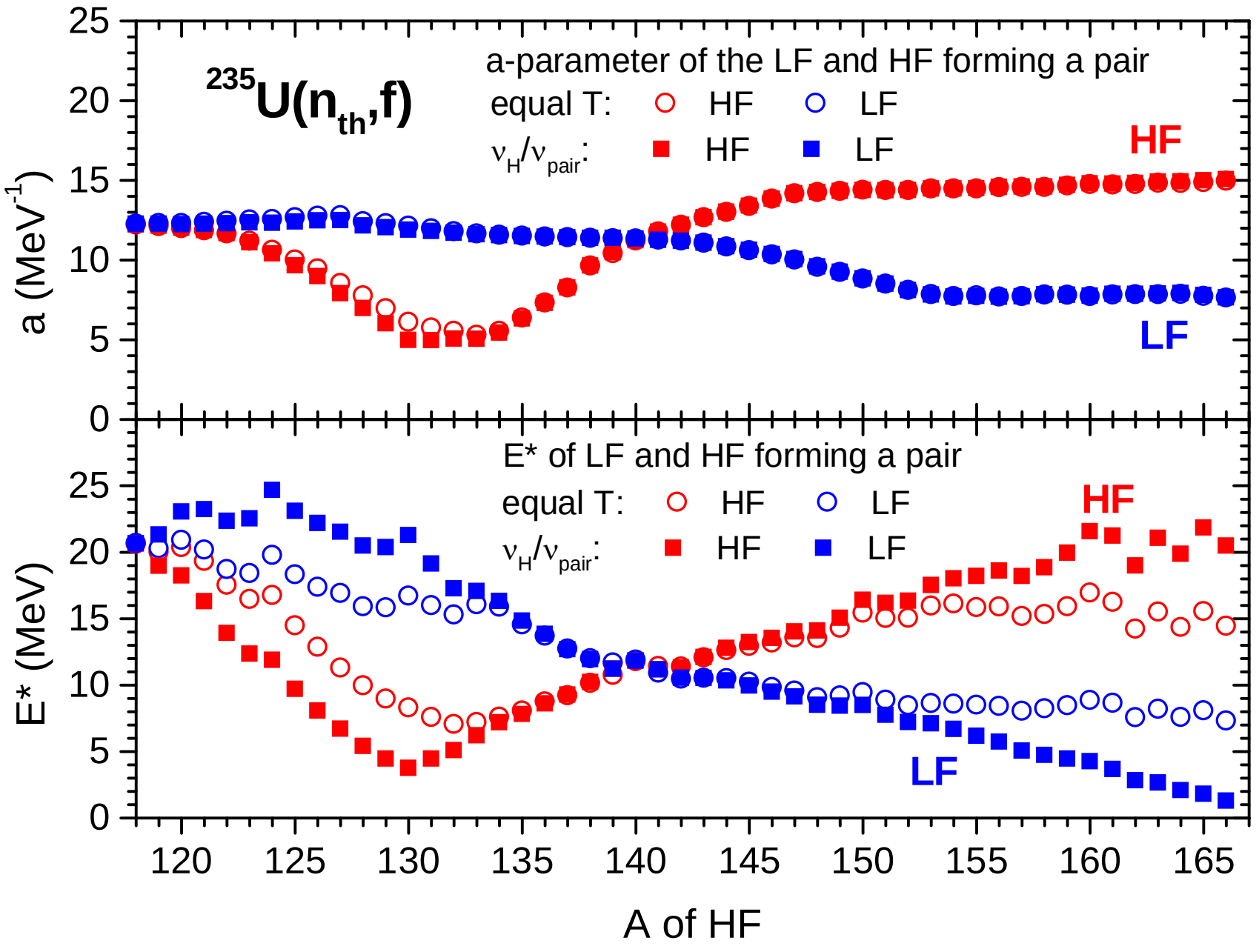

**Fig.4**: $^{235}U(n_{th},f)$ E* and a-parameter of HF (red symbols) and LF (blue symbols) forming a pair as a function of $A_H$, a-parameters obtained from the $\nu_H/\nu_{pair}$ parameterization with full squares and from the "equal T" method with open circles.

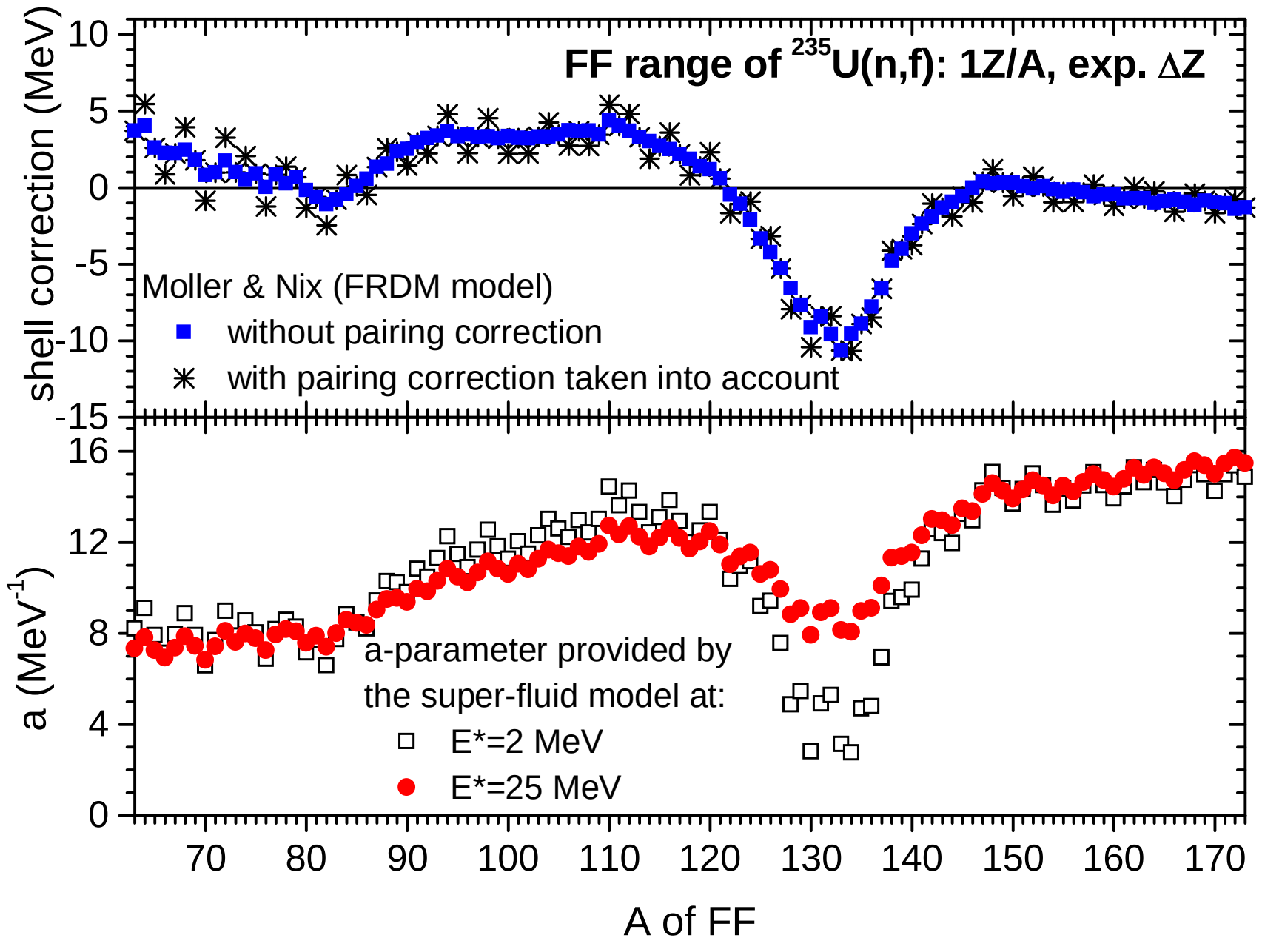

**Fig.5**: Upper part: shell corrections of the FF range of $^{235}U(n,f)$ with (stars) and without (squares) pairing corrections. Lower part: a-parameter (super-fluid model) calculated at E*=2 MeV (open squares) and E*=25 MeV (full circles).

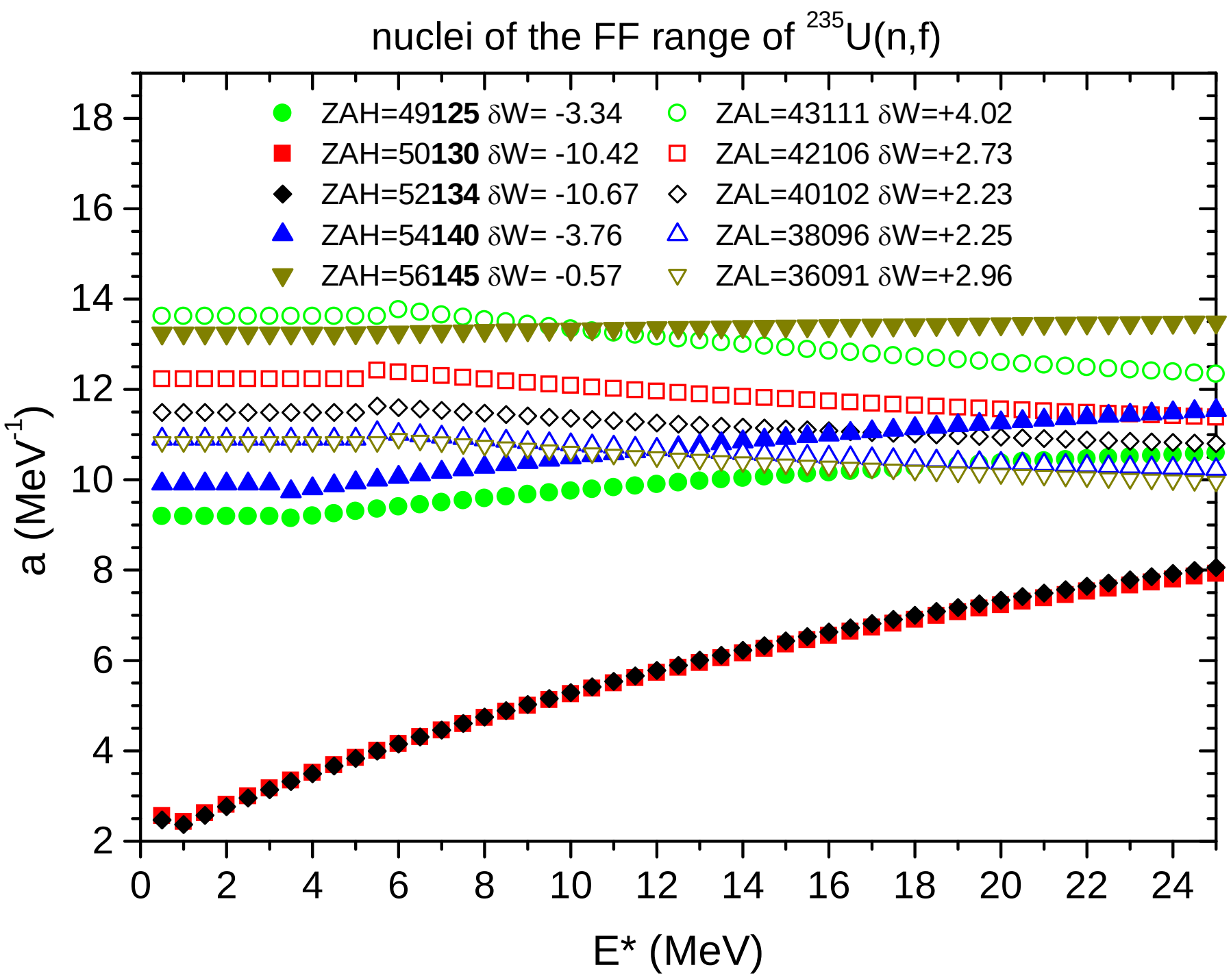

**Fig.6**: a-parameter variation with E* illustrated for 5 pairs of the FF range of $^{235}$U(n,f), a-parameters of HF with different full symbols and of the complementary LF with the corresponding open symbols.

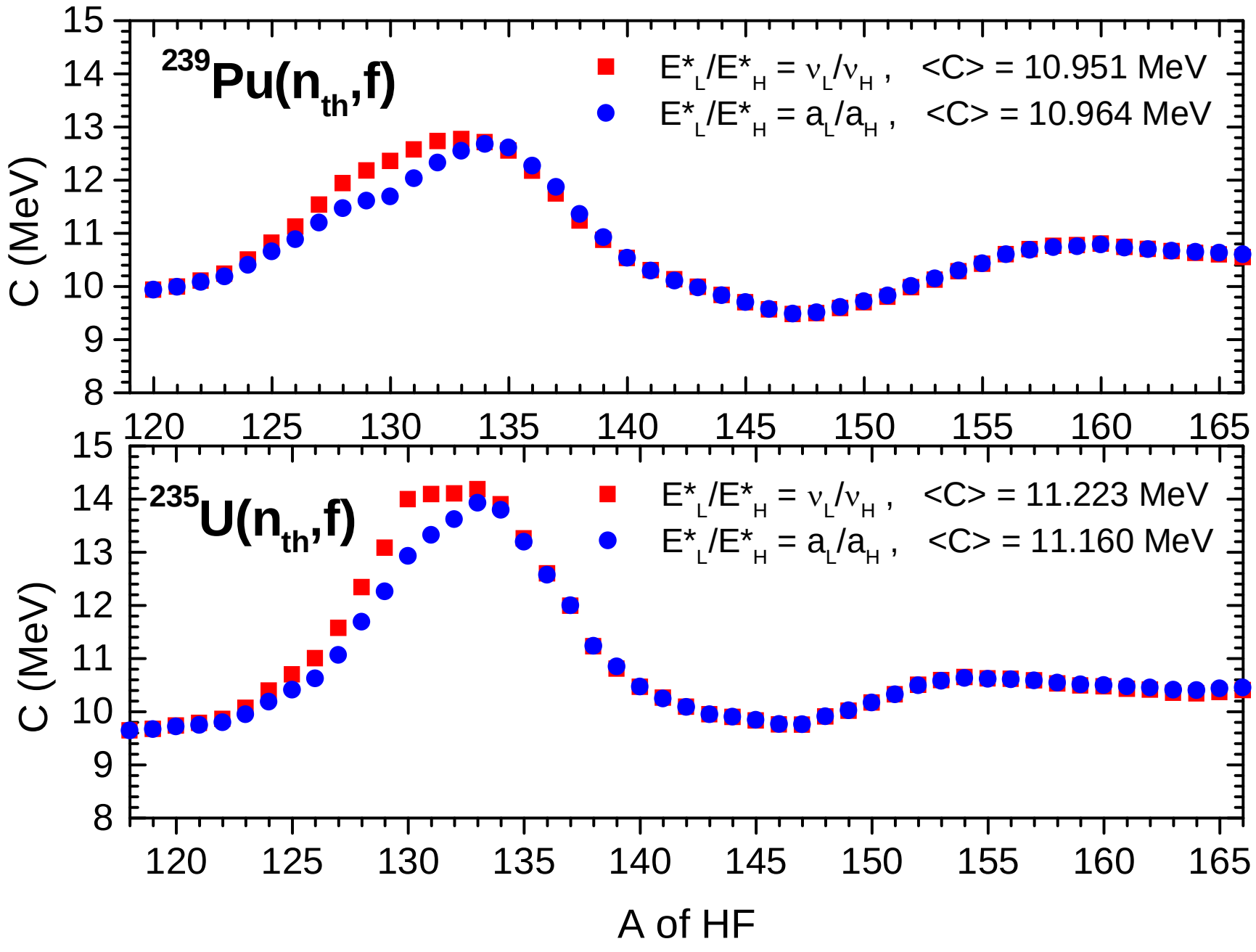

**Fig.7**: C-parameter of the FF pairs versus $A_H$ obtained by the $\nu_H/\nu_{pair}$ parameterization (full red squares) and “equal T” methods (full blue circles) for $^{239}$Pu($n_{th}$,f) (upper part) and $^{235}$U($n_{th}$,f) (lower part).

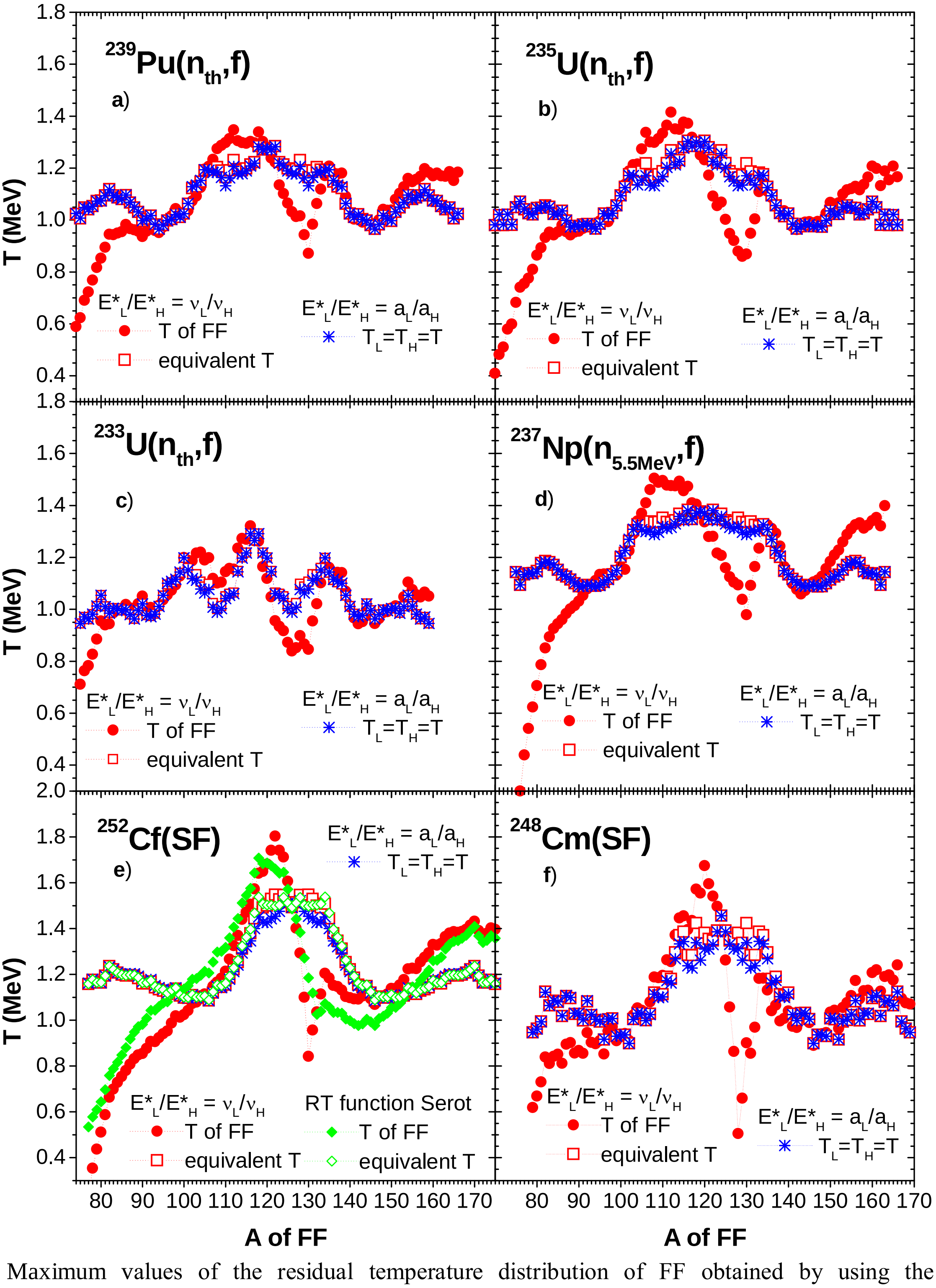

**Fig.8**: Maximum values of the residual temperature distribution of FF obtained by using the $\nu_H/\nu_{pair}$ parameterization (fragment temperature with full circles, and the equivalent temperature with open squares) in comparison with the "equal T" method (with stars). In addition for $^{252}$Cf(SF) the temperatures from the RT function of Serot are plotted with green diamonds.

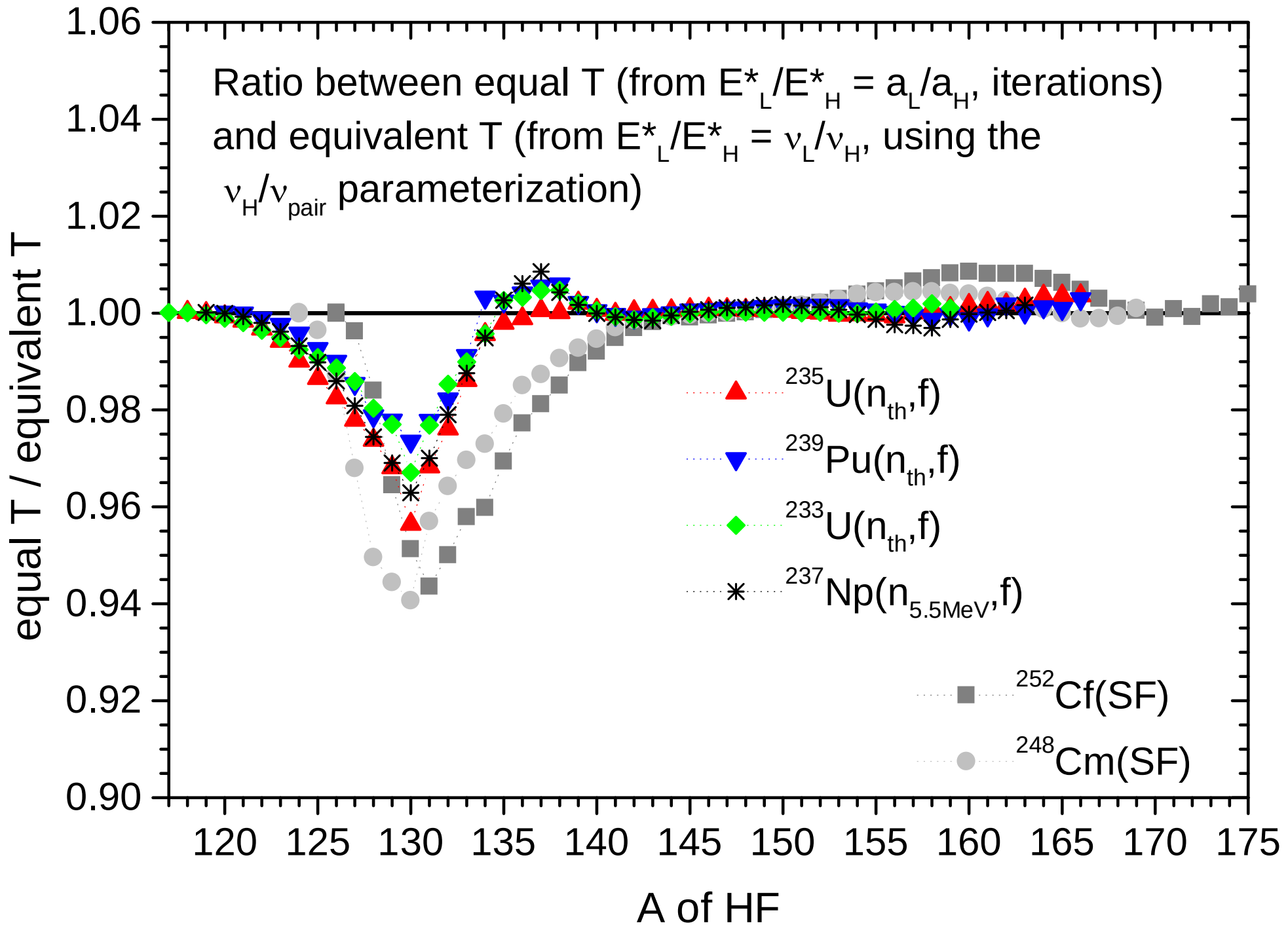

**Fig.9:** Ratio between equal residual temperature and equivalent residual temperature (from the $\nu_H/\nu_{pair}$ parameterization).

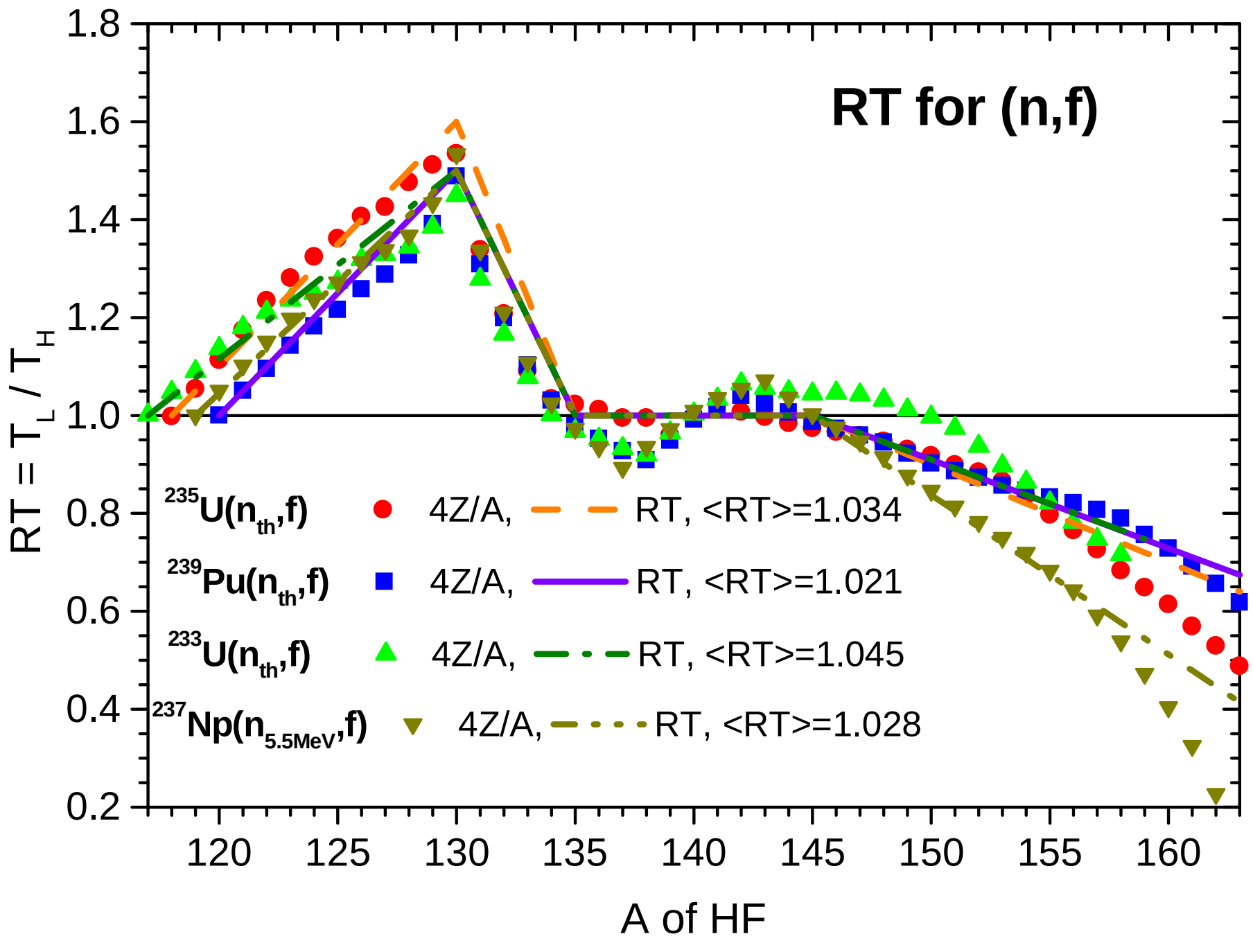

**Fig. 10**. FF temperature ratios and parameterizations for neutron induced fission.

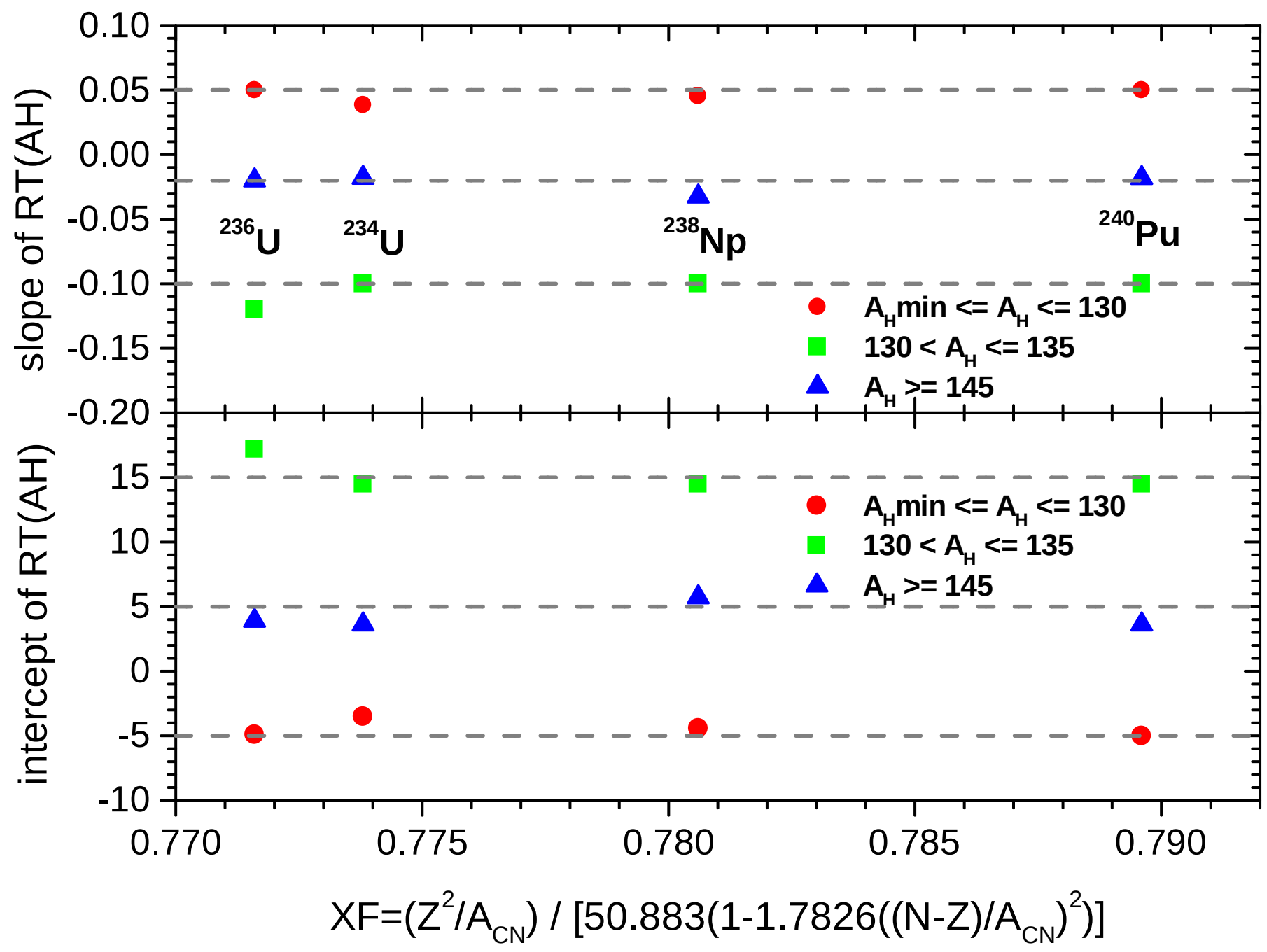

**Fig. 11**. Slopes and intercepts of the RT parameterizations for neutron induced fission.

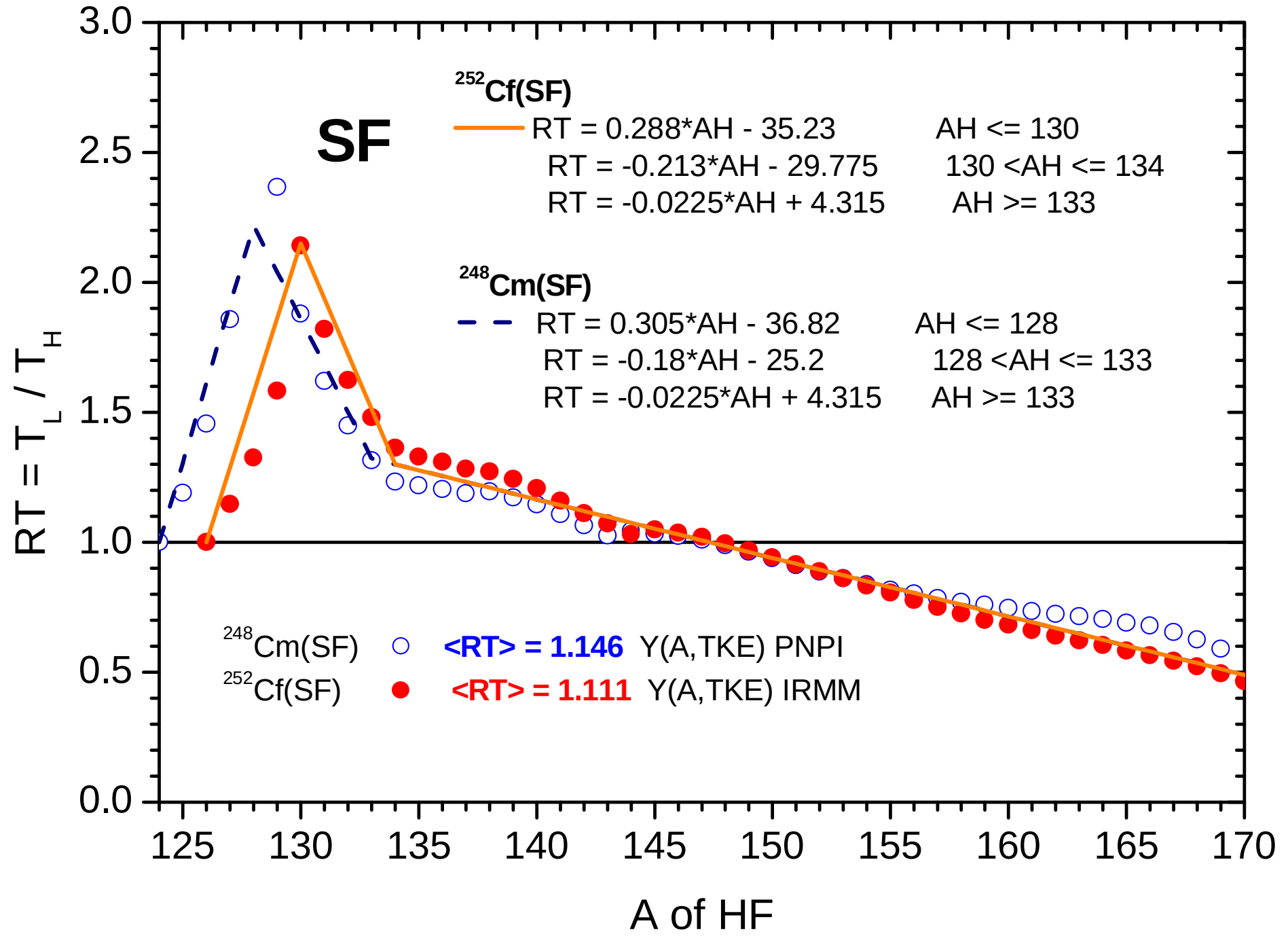

**Fig. 12**. FF temperature ratios and parameterizations for the spontaneous fissioning systems.

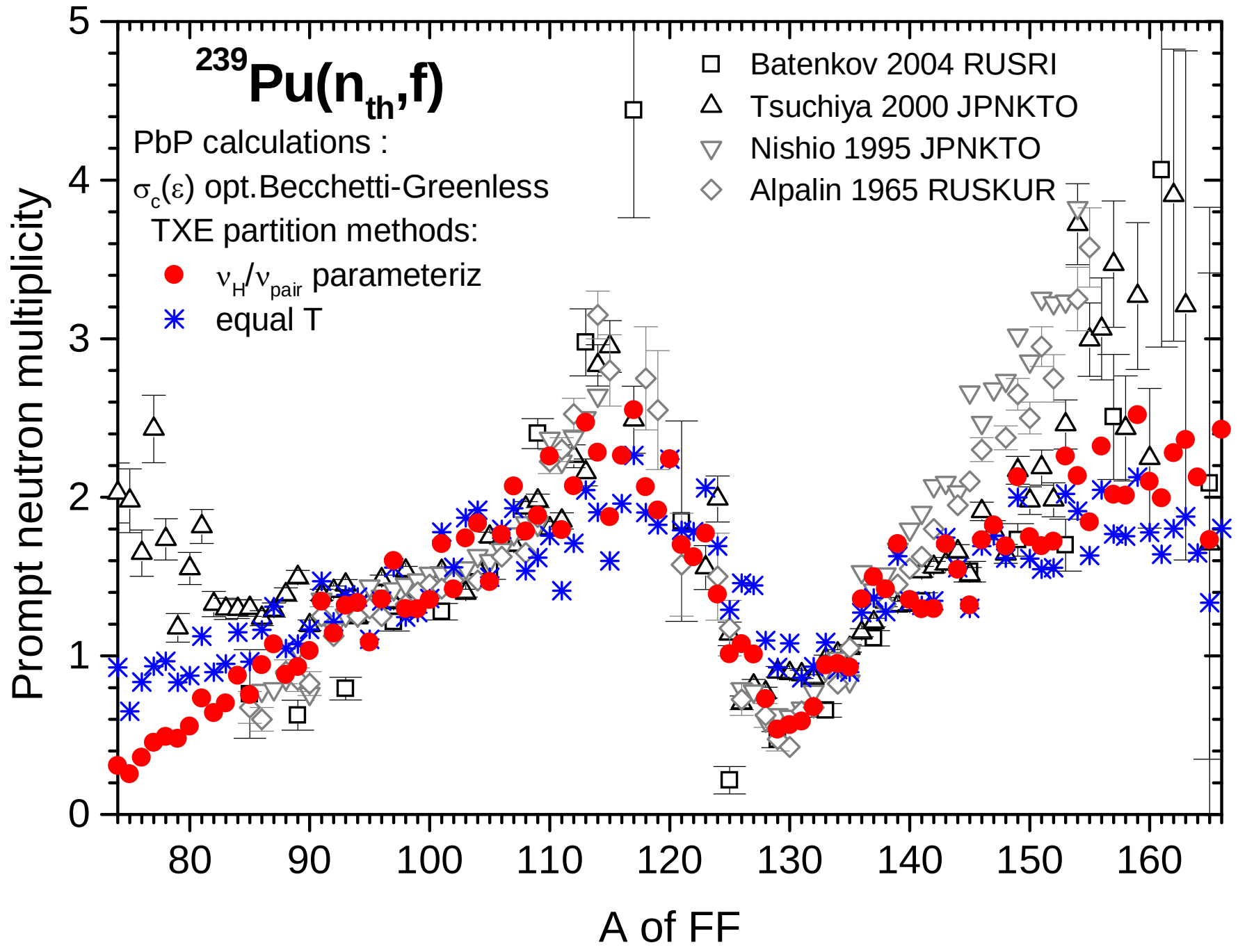

**Fig.13**. $^{239}$Pu(n$_{th}$,f) ν(A) calculations using the two TXE partition methods, in comparison with experimental data from EXFOR.

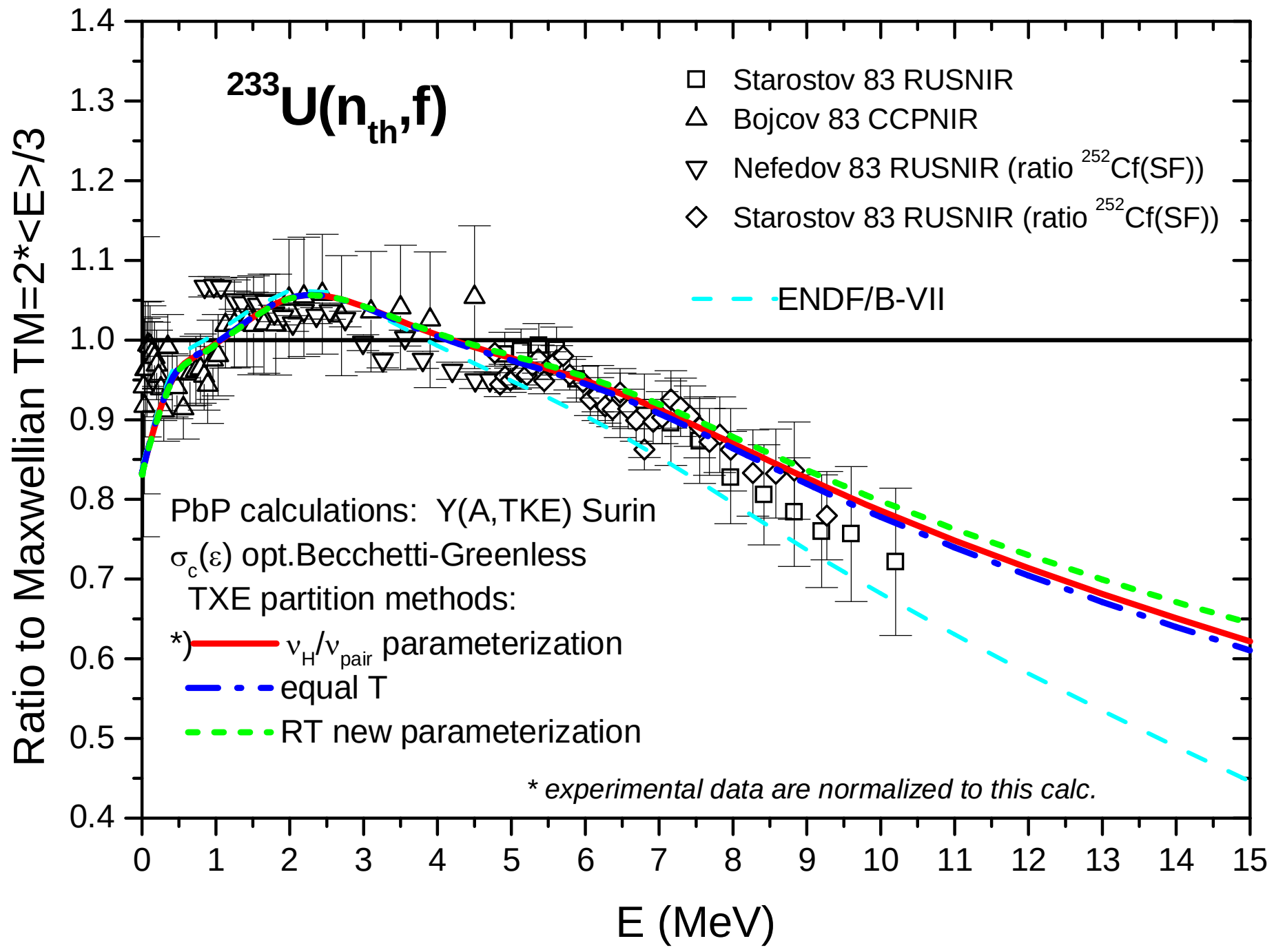

**Fig.14**. $^{233}$U(n$_{th}$,f) Prompt fission neutron spectrum calculations (given as ratio to a Maxwellian spectrum) using the two TXE partition methods and the RT parameterization, in comparison with experimental data from EXFOR and with the ENDF/B-VII evaluation.

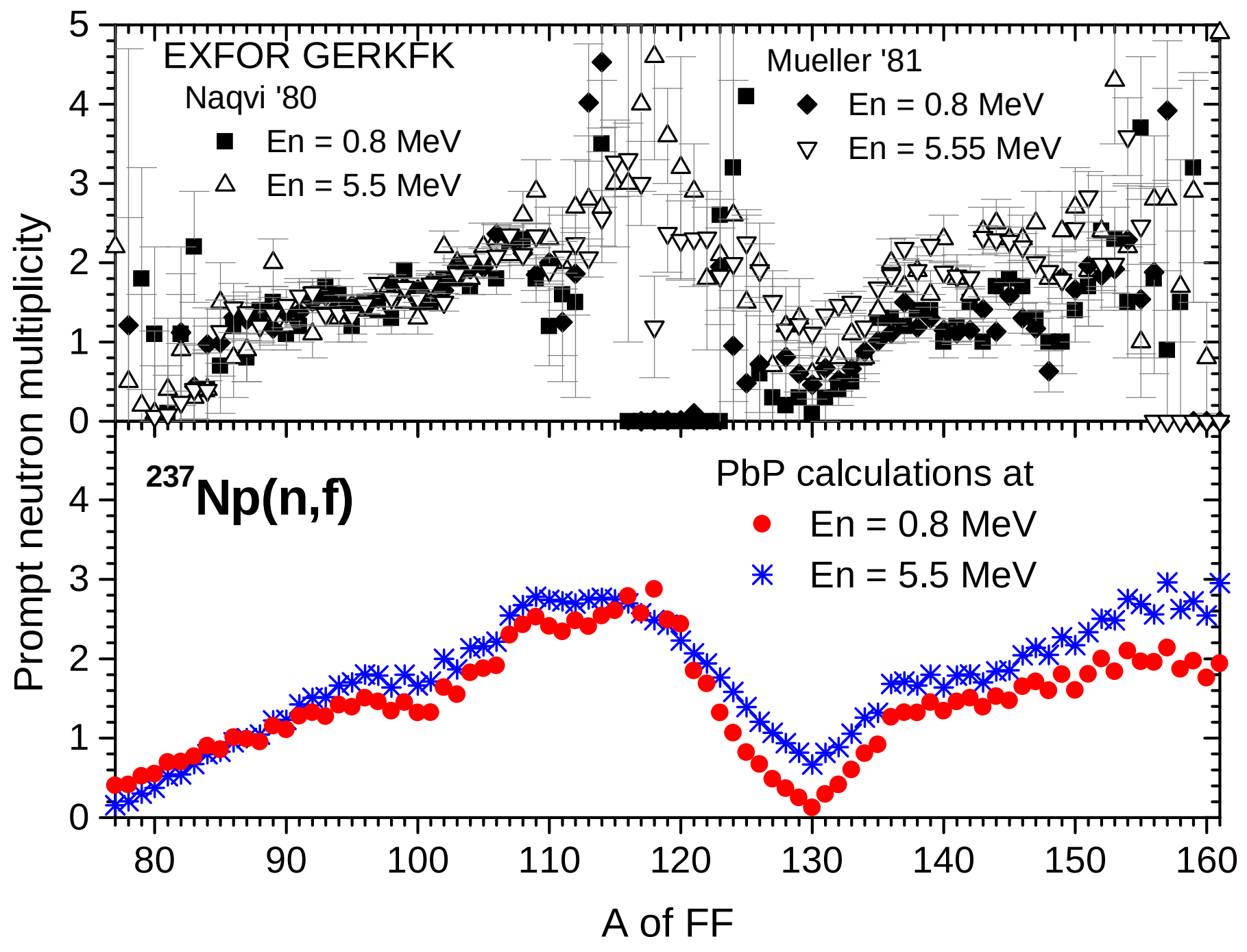

**Fig.15**: ν(A) of $^{237}$Np(n,f), upper part: the experimental data sets of Naqvi and Mueller at En=0.8 MeV (full symbols) and En=5.5 MeV (open symbols). Lower part: PbP model calculation at En=0.8 MeV (full circles) and En=5.5 MeV (stars).

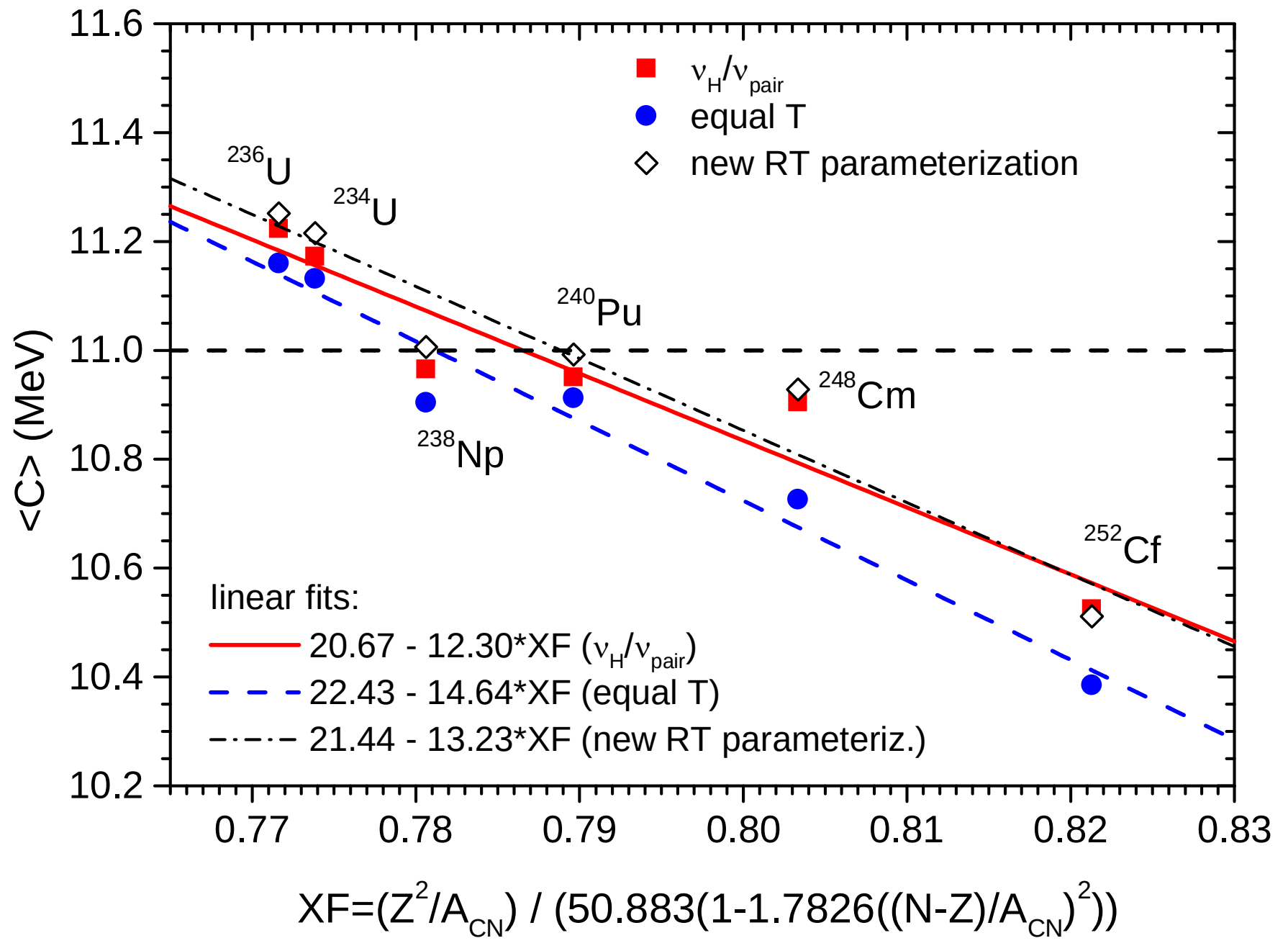

**Fig.16**: Average value of the C parameter as a function of the fissility parameter.

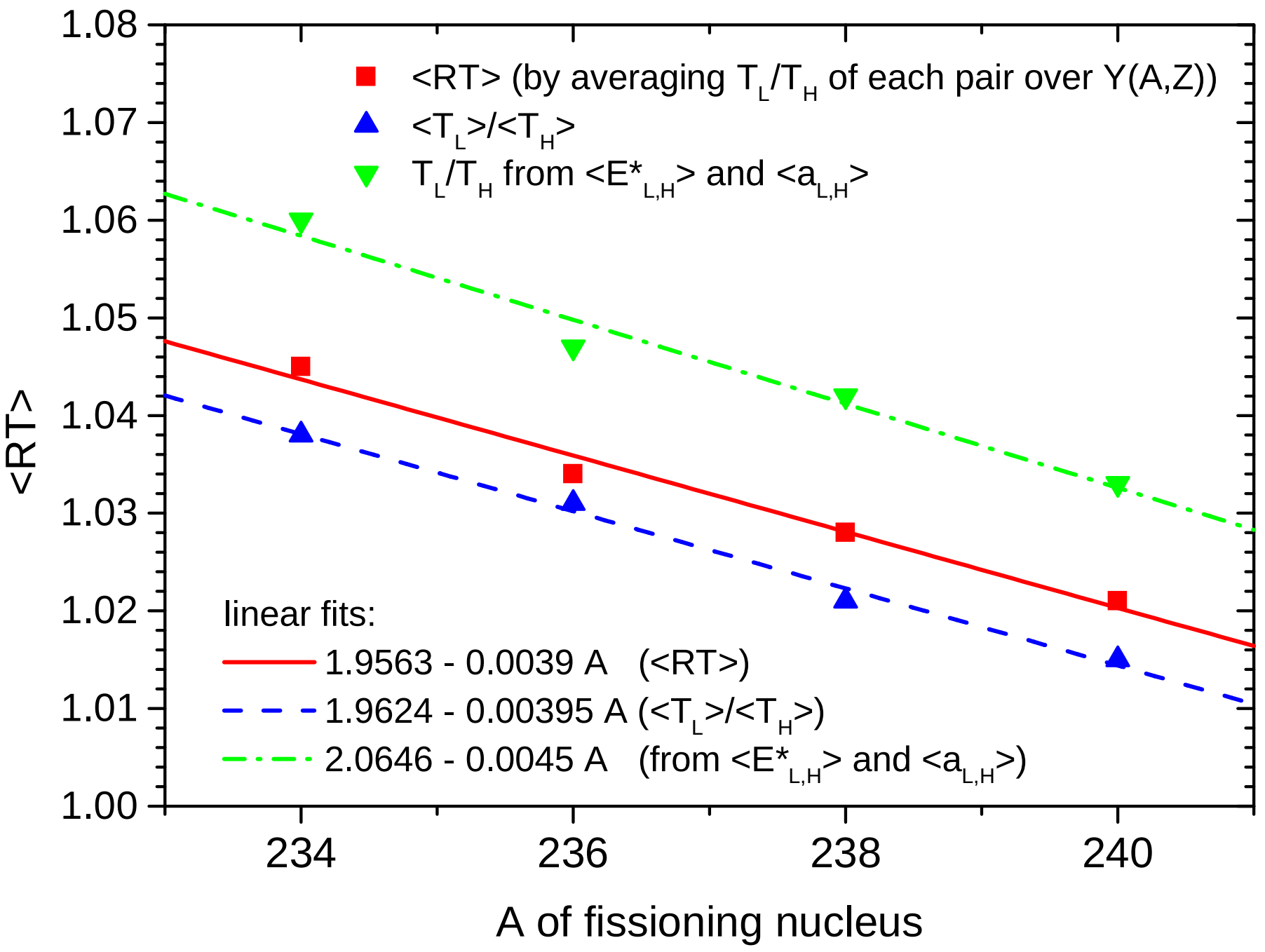

**Fig.17**: Average value of the fragment temperature ratios as a function of the mass number of the fissioning nucleus.